\documentclass[journal]{IEEEtran}
\usepackage{multicol}
\usepackage{multirow}

\usepackage{amsmath}
\usepackage{subfigure}
\usepackage{amsthm}
\usepackage{setspace} 
\usepackage{psfrag}
\usepackage{amsfonts}
\usepackage{amssymb}
\usepackage{stmaryrd}
\usepackage{mathrsfs}
\usepackage{epstopdf}
\usepackage{enumerate}
\usepackage{booktabs}  
\usepackage{amsmath, color}
\usepackage{dsfont,bm}
\usepackage{amssymb}
\usepackage{epsfig}
\usepackage{graphicx}
\usepackage{amssymb}
\usepackage{epsfig}
\usepackage{epstopdf}
\usepackage{subfigure}
\usepackage{balance}
\usepackage{multicol}
\usepackage{enumerate}
\usepackage{verbatim,algorithm,algorithmic,cite}
\usepackage{booktabs}
\usepackage{url}
\usepackage{cases}
\usepackage{stmaryrd}

\usepackage[inline]{enumitem}
\ifCLASSINFOpdf
\else
\fi
\makeatletter

\newcommand{\Rmnum}[1]{\expandafter\@slowromancap\romannumeral #1@}
\makeatother

\begin{document}
%

\title{Double-Phase-Shifter Based Hybrid Beamforming for mmWave DFRC in the Presence of Extended Target and Clutters}

\author{\large Ziyang~Cheng,~\IEEEmembership{Member~IEEE,} Linlong~Wu,~\IEEEmembership{Member~IEEE,} Bowen~Wang,~\IEEEmembership{Student  Member~IEEE,} Bhavani~Shankar~M.~R.,~\IEEEmembership{Senior Member~IEEE,} and~Bj\"{o}rn~Ottersten,~\IEEEmembership{Fellow~IEEE}
\thanks{The work of Ziyang Cheng and Bowen Wang was supported in part by the National Natural Science Foundation of China under Grants 62001084 and 62031007, and in part by the National Defense Science and Technology Foundation under Grant 2022--JCJQ--JJ--0202. 
The work of Linlong Wu, Bhavani~Shankar and Bj\"{o}rn Ottersten was supported in part by ERC AGNOSTIC under Grant EC/H2020/ERC2016ADG/742648, and in part by FNR CORE SPRINGER under Grant C18/IS/12734677. \textit{(Corresponding author: Linlong Wu)}}
\thanks{
Ziyang Cheng and Bowen Wang are with the School of Information \& Communication Engineering, University of Electronic Science and Technology of China, Chengdu 611731, China. Email: {zycheng}@uestc.edu.cn, B\_W\_Wang@163.com.}
\thanks{
Linlong Wu, Bhavani~Shankar and Bj\"{o}rn Ottersten are with the Interdisciplinary Centre for Security, Reliability and Trust (SnT), University of Luxembourg, Luxembourg City L-1855, Luxembourg. Email: \{linlong.wu, bhavani.shankar, bjorn.ottersten\}@uni.lu.}
\thanks{The conference precursor of this work was presented in the 2022 European Signal Processing Conference (EUSIPCO) \cite{Wang2022Hybrid}.}
}

\maketitle

\begin{abstract}
In millimeter-wave (mmWave) dual-function radar-communication (DFRC) systems, hybrid beamforming (HBF) is recognized as a promising technique utilizing a limited number of radio frequency chains. 
In this work, in the presence of extended target and clutters, a HBF design based on the subarray connection architecture is proposed for a multiple-input multiple-output (MIMO) DFRC system. In this HBF, the double-phase-shifter (DPS) structure is embedded to further increase the design flexibility. 
We derive the communication spectral efficiency (SE) and radar signal-to-interference-plus-noise-ratio (SINR) with respect to the transmit HBF and radar receiver, and formulate the HBF design problem as the SE maximization subjecting to the radar SINR and power constraints.
To solve the formulated nonconvex problem, the join{T} {H}ybrid b{E}amforming and {R}adar r{E}ceiver {O}ptimizatio{N} (THEREON) is proposed, in which the radar receiver is optimized via the generalized eigenvalue decomposition, and the transmit HBF is updated with low complexity in a parallel manner using the consensus alternating direction method of multipliers (consensus-ADMM). 
Furthermore, we extend the proposed method to the multi-user multiple-input single-output (MU-MISO) scenario.
Numerical simulations demonstrate the efficacy of the proposed algorithm and show that the solution provides a good trade-off between number of phase shifters and performance gain of the DPS HBF.
\end{abstract}



\IEEEpeerreviewmaketitle


%
\IEEEpeerreviewmaketitle
\vspace{-0.6em}
\section{Introduction}
\IEEEPARstart{F}{uture} 6th Generation (6G)  mobile communication systems are expected to  possess a sensing capability to enable various connected service applications\cite{8869705}, such as unmanned aerial vehicles (UAVs) and intelligent automobiles \cite{7786130}. Such applications require larger amounts of spectrum, which makes it unaffordable to assign independent bands to the radio-frequency (RF) systems. Therefore, integrated sensing and communications (ISAC), as a technology with improved spectrum efficiency, lower power consumption and reduced cost, will play a crucial role in 6G and beyond \cite{liu2020joint}.

The approaches to ISAC so far can be roughly categorized into two groups, namely, co-existence and dual function of radar and communication. For the group of the co-existence of radar and communication \cite{zheng2019radar,mishra2019toward}, the two systems operate with independent transmitters sharing the same frequency band. Although this approach also improves the spectral
efficiency, it could suffer from the inevitable mutual interference between radar and communication, which is in fact the key research issue in the related literature. The straightforward way is to design a spectrally compatible waveform (SCW) \cite{nunn2012spectrally,aubry2014radar,aubry2016forcing,wu2017transmit,cheng2018spectrally,8770133}.  Such approaches require to sense the spectrum occupied  by the communication, and design radar waveforms with desired spectrum nulls to avoid imposing interference produced by radar on the communication.  Although the SCW can be implemented to guarantee the  co-existence, it does not really achieve the communication and radar spectrum sharing (CRSS) in a true sense given that the radars only operate at the frequency bands which are unoccupied by communications.  Therefore,  many co-design methods \cite{sodagari2012projection,li2016optimum,li2017joint,liu2018mu,mahal2017spectral,zheng2017joint, cheng2019codesign} were proposed to overcome the limitations of the SCW methods. The pioneering work on the co-design for the co-existence  of radar and communication was proposed in \cite{li2016optimum}, where  a co-design of communication covariance matrix and radar sub-sampling matrix is proposed to minimize the interference caused to radar keeping the  constraints of power and capacity for achieving the co-existence of  matrix completion multiple-input multiple-output (MC-MIMO) radar and MIMO communication system. Moreover, the authors considered the joint design of radar waveform and communication precoding matrix  for the co-existence scenario under the signal-dependent clutter environment in \cite{li2017joint}. In addition, the co-existence of a communication system and pulsed radar and in the presence of signal-dependent interference was considered in \cite{zheng2017joint}, where the radar pulse codes and communication precoding matrix are jointly optimized to maximize the compound rate while guaranteeing the constraints of power and radar signal-to-interference-plus-noise-ratio (SINR). 

In contrast to the studies on the co-existence, the second group aims to build a dual-function radar-communication (DFRC) system \cite{chiriyath2015inner}, where the communication and radar sensing functions are integrated into one platform and thereby allows a dual-function waveform to achieve both sensing and communication simultaneously. 
Such DFRC systems transmit dual-function waveforms by considering both radar and communication performance metrics jointly, and have gained a growing attention recently \cite{blunt2010embedding, blunt2011performance,blunt2010intrapulse, hassanien2015dual,hassanien2016phase,hassanien2017dual,dokhanchi2019mmwave, cheng2021transmit,dokhanchi2021adaptive,liu2018toward,8445927}. 
For example, as a simple way to achieve DFRC, the beampattern of a MIMO radar is optimized to implement traditional communication modulations, such as phase shift keying (PSK) and amplitude shift keying (ASK), by controlling sidelobe level of MIMO radar beampattern \cite{hassanien2015dual,hassanien2017dual}. In addition to these single-carrier methods, the orthogonal frequency division multiplexing (OFDM) signal is regarded  as a promising candidate for the DFRC waveform \cite{5483108}. In \cite{5776640},  the OFDM-based method  employs the  fast Fourier transform (FFT) and the inverse FFT (IFFT) to obtain the Doppler and  range parameters, respectively. Besides, \cite{9420261} designed a time-frequency waveform  for an OFDM  DFRC system which communicates with an OFDM receiver while
estimating target parameters simultaneously. Apart from these, in \cite{liu2018toward}, the authors proposed several beamforming designs to implement a joint MIMO radar transmission and MU-MIMO communication by shaping a desired radar beampattern while keeping the downlink SINR and power requirements.

However, the existing methods developed for the DFRC system implicitly assume a fully-digital architecture, in which an independent radio frequency (RF) chain is associated with each antenna including a mixer and a digital-to-analog converter (DAC) or analog-to-digital converter (ADC). This architecture might lead to extremely high hardware costs and power consumption, especially for large-scale millimeter-wave (mmWave) systems. As a result, the hybrid beamforming (HBF) architecture is viewed as a practical solution to the DFRC system. Specifically, in the HBF structure, a small number of RF chains are needed to ensure the satisfactory performance and large number of RF phase shifters (PSs) are adopted to reduce the cost. For communication-only systems, the HBF has been fully developed for both single-user (SU) and multi-user (MU) scenarios \cite{zhang2014achieving,yu2016alternating,han2015large,sohrabi2016hybrid,wang2018hybrid,6717211,7448873,mo2017hybrid}, but it has been less studied for DFRC systems. In fact, HBF has been proposed for DFRC systems for the first time in \cite{8683591}, where the HBF is optimized to approach the the performance of ideal digital beamformer by considering the weighted summation of the radar and communication beamforming errors. However, the work in \cite{8683591} is based on the "two-stage" approach, i.e., the ideal digital radar and communication beamformers are firstly obtained, the HBF is then optimized according to the ideal digital beamformers. This indirect design procedure may not guarantee a satisfactory performance or exploit the systems full potential. Towards that end, the DFRC HBF with subarray-connection structure is considered in \cite{cheng2021hybrid}, where the HBF is designed to maximize the sum-rate subject to power and HBF constraints.  
 
In terms of system model, the above two works do not consider the signal-dependent interference (such as clutters) environment, which is usually considered as the main challenge for sensing the target in radar applications \cite{wu2017transmit,wu2016cognitive}. 
{\color{black}Moreover, for 
large-size target and clutter, their echoes   become extended  over range cells} \cite{1705013,4840496,4200705}. 
Different from the point-like target scenario, in the presence of extended target and clutters, the design requires some prior knowledge of the target and clutters, such as their impulse response or their statistics. 
Consequently, the model of the extended target and clutters is more complicated. 
To the best of our knowledge, the HBF design has not been investigated for the DFRC system in environments with extended target and clutters in the literature. In addition, the conventional single-phase-shifter (SPS) structure reduces the hardware cost of a mmWave system while bearing a performance loss.
\textcolor{black}{
The DPS structure \cite{bogale2016number,lin2016quantization,yu2019doubling}, where each antenna is connected to two in-parallel phase shifters, has been widely investigated in the communication field.
For example, the authors in \cite{bogale2016number} propose zero-forcing (ZF) based heuristic algorithms to select antennas and optimize DPSs jointly.
As a further step, a two-stage  algorithm for designing DPS-based HBF is investigated in \cite{yu2019doubling}.
The corresponding results show that exploiting DPS can achieve a balanced trade-off between performance and cost.
However, the above works focus on the communication-only systems, and the  methods are difficult to extend to   DFRC systems.
}
Motivated by these facts, in this paper, we investigate the HBF design problem based on the DPS structure for the mmWave DFRC system in the presence of extended target and clutters. 
The main contributions of this work are summarized as follows:
\begin{itemize}
\item We propose a novel hardware architecture for the analog beamforming component of the HBF based DFRC system, which adopts a DPS structure associated with each antenna. Compared with the conventional SPS structure, the DPS provides an extra degree of freedom (DoF) (i.e. amplitude control) of design.
To adapt to the proposed HBF architecture, we derive the corresponding communication spectral efficiency (SE) and radar SINR as performance metrics and then formulate the DPS-based HBF design problem. 
\item  An algorithm termed join{T} {H}ybrid b{E}amforming and {R}adar r{E}ceiver  {O}ptimizatio{N} (THEREON) is proposed to solve the formulated nonconvex problem. 
For the radar receive filter, it is updated via the generalized eigenvalue decomposition. For the HBF design, the weighted minimum mean-square error (WMMSE) reformulation \cite{shi2011iteratively} is adopted first, and then we solve the corresponding problem based on the consensus alternating direction method of multipliers (consensus-ADMM) \cite{boyd2011distributed}, in which the closed form solutions of the primal variables are derived via the Karush-Kuhn-Tucker (KKT) conditions. In addition, the proposed algorithm is adapted to the multi-user multiple-input single-output (MU-MISO) scenario. 
\item Representative simulations are conducted to illustrate the efficacy of the proposed algorithm and the performance improvement enabled by the proposed DPS architecture. For different initialization, the algorithm consistently converges to a point with improved SE. We also demonstrate that the DPS structure can substantially improve the performance comparing to the conventional SPS structure at the cost of only an extra phase shifter for each transmit antenna. 
\end{itemize}

The remainder of the paper is organized as follows. In Section II, the signal model and problem formulation are presented. The proposed algorithm is developed in Section III, and then extended to the MU-MISO scenario in Section IV. Section V presents various numerical simulations. Conclusions are drawn  in Section VI.

\textit{Notation:} Lower case  and upper case bold face letters denote  vectors  and   matrices, respectively.    $ (\cdot)^* $,   $ (\cdot)^H $ and  $ (\cdot)^T $    represent the conjugate,    conjugate transpose  and transpose operators, respectively. $\mathbb{C}^{n}$  and $\mathbb{C}^{N\times N}$   denote the sets of $n$-dimensional complex-valued   vectors and $ N \times N $ complex-valued   matrices, respectively. The  real part   of a complex-valued number  and expectation operator  are noted by  $ \Re \left\{  \cdot  \right\} $ and $  {\mathbb E}\left\{   \cdot \right\} $. 
 ${\rm Tr}({\bf A})$ is reserved for  the trace of $ \bf A $.    $ \mathbf{I}_{N} $ denotes the $ {N\times N} $ identity matrix.     
 $\|\cdot\|_F $ and $\jmath = \sqrt{-1}$ denote the Frobenius norm and the imaginary unit,  respectively. 
 Finally, ${\rm Bdiag}(\cdot)$ and ${\rm diag}(\cdot)$ stand for the block diagonal matrix and the vector composed by the diagonal entries of a matrix, respectively.

\vspace{-1em}
\section{Signal Model and Problem Formulation}
\begin{figure}[!t]
	\centering
	\subfigure[]
	{\includegraphics[scale=0.5]{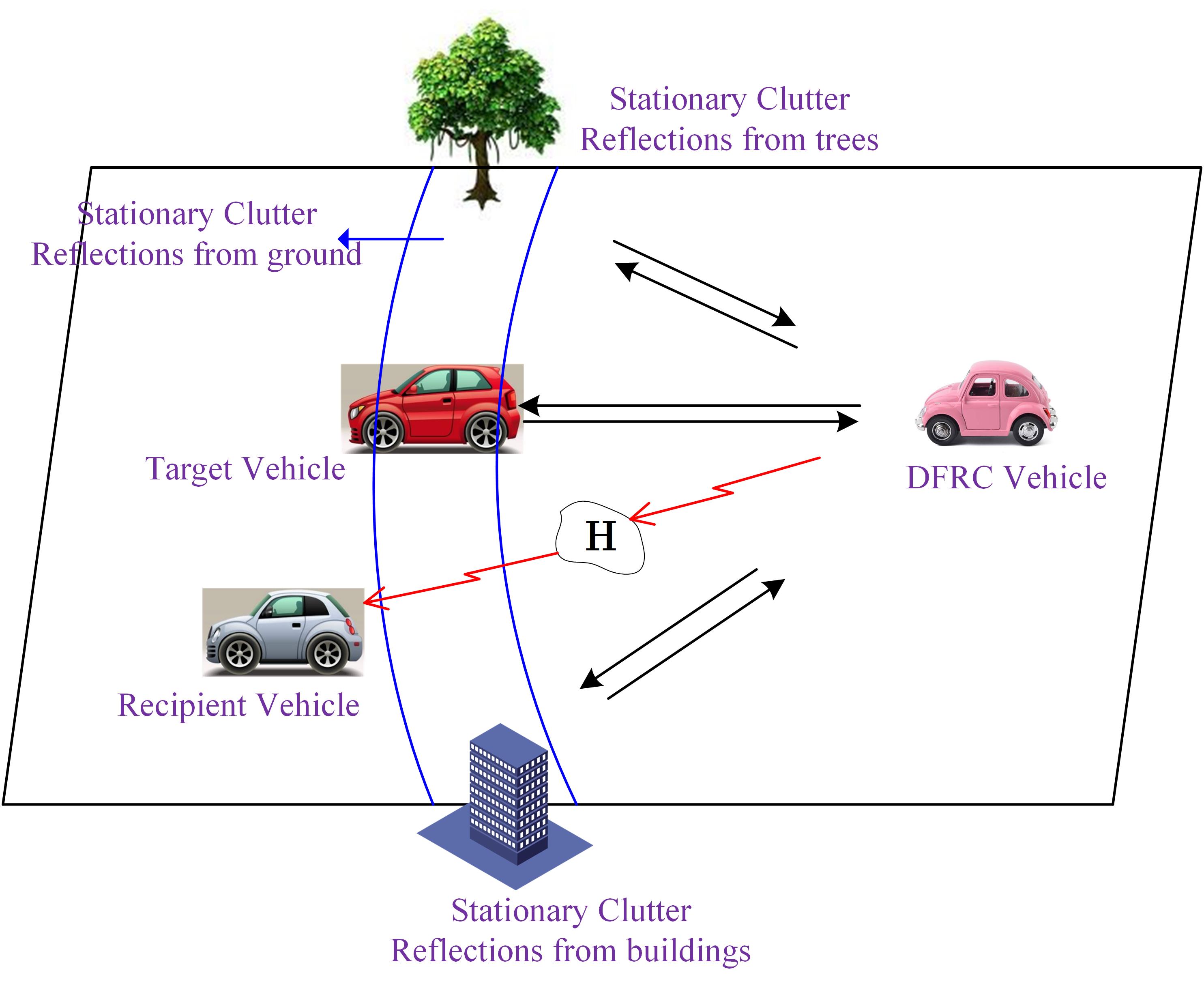}}
    \subfigure[]
    {
    \includegraphics[scale=0.65]{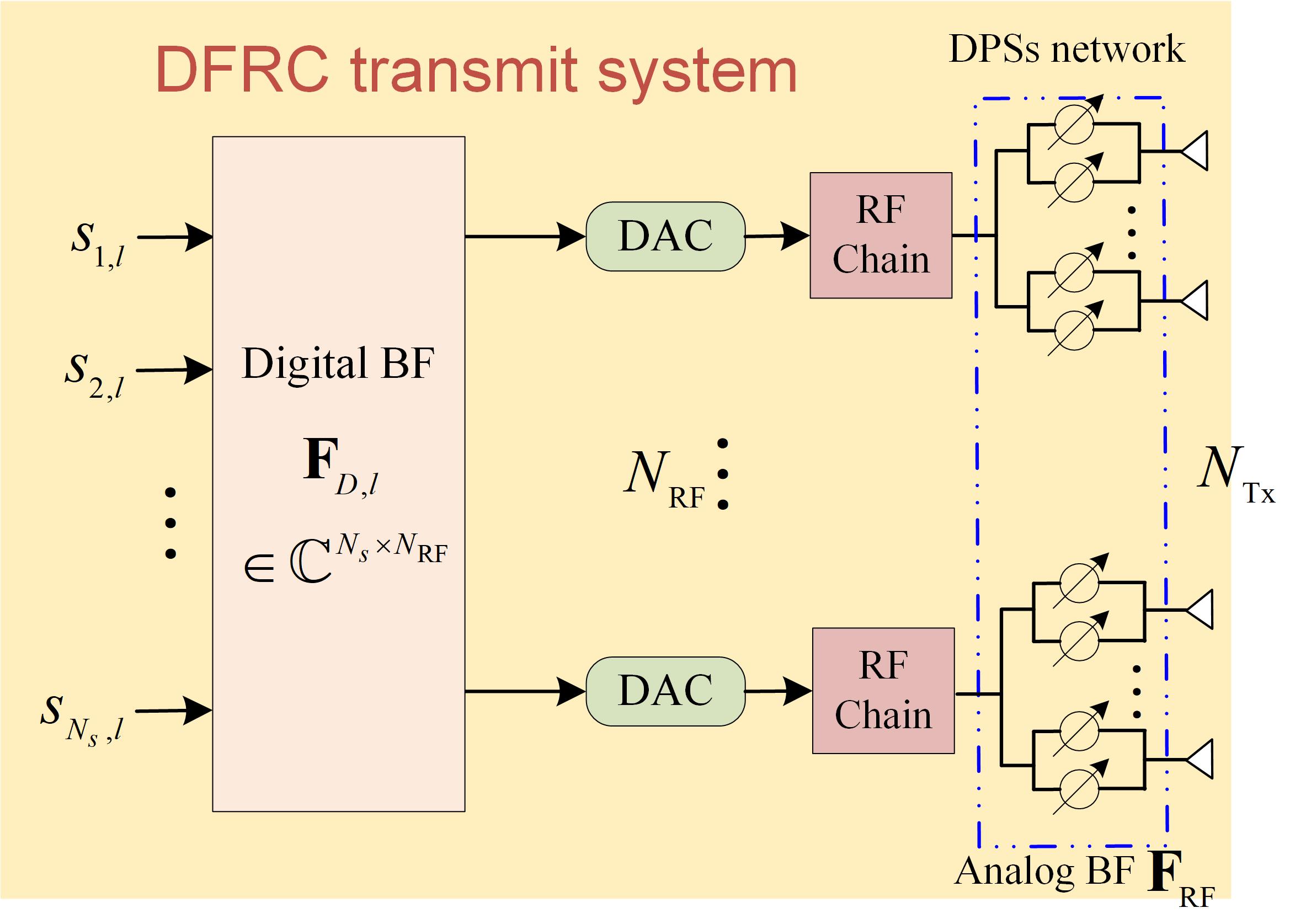}
}
	\caption{(a) Illustration of our considered scenario for the DFRC system.
	(b) Overview of the DPS-based HBF DFRC system.}
	\label{fig:pic1}
\end{figure}

In this section, we formulate the system model  and optimization problem for the proposed  HBF DFRC system. We consider a  scenario as shown in Fig. \ref{fig:pic1}(a), where a DFRC vehicle sends communication symbols to a recipient vehicle receiver while detecting a target vehicle of interest in the presence of stationary clutters (such as trees, ground, buildings, etc.) simultaneously.  The system architecture is depicted in Fig. \ref{fig:pic1}(b), where   we assume 
   a  time-division duplex (TDD) DFRC system with $N_{\rm Tx}$ antennas and $ N_{\rm RF} $ RF chains adopting a   non-overlapping subarray architecture. Each  subarray has $M= N_{\rm Tx}/N_{\rm RF}$
antennas connected to a  RF chain.   The   recipient vehicle receiver  with $N_{\rm Rx}$ antennas employs the fully-digital beamforming structure.  

\subsection{Transmit Model}
At the transmitter, the symbol block ${\bf s}_l$ in $l$-th subpulse\footnote{ {We use the term subpluse and slot exchangeably in this paper, where the former is commonly used in radar and the later is used in communications.}} is   precoded, at first, by a digital precoding matrix ${\bf F}_{{\rm D}, l} \in {\mathbb C}^{N_{\rm RF} \times N_s}$, where $N_s$ is the number of data streams. Subsequently, the  baseband signal is up-converted to the RF domain via $N_{\rm RF} $ RF chains and processed by analog PSs. Different from the conventional subarray architecture where each antenna is connected to a single PS, we consider exploiting double PSs to provide additional amplitude control 
for the HBF, the diagram of which is sketched in Fig. \ref{fig:pic1}(b).

Without loss of generality, each PS has a constant magnitude $ \frac{1}{\sqrt{N_{\rm Tx}}} $, and the synthesized value of each DPS module meets $Ae^{\jmath \varphi}$ with $A\in [0, {2}/{\sqrt{N_{\rm Tx}}}]$
and $\varphi \in [0, 2\pi]$.  Thus, the proposed analog precoder can be expressed as 
\begin{equation}
 {\bf F}_{\rm RF}= {\bf F}_{\rm set} {\bf P},
\end{equation}
where $ {\bf F}_{\rm set} ={\rm diag}\{ {f}_{1},\cdots,    {f}_{N_{\rm Tx}} \}$ with $ {f}_m=A_{m}e^{\jmath \varphi_{m}}$,  $A_{m}\in [0, {2}/{\sqrt{N_{\rm Tx}}}]$,$\varphi_{m} \in [0, 2\pi]$, $\forall m=1,\cdots,N_{\rm Tx}$,
$ {\bf P}= {\rm Bdiag}  \left\lbrace {\bf 1}_{M}, \cdots, {\bf 1}_{M} \right\rbrace \in {\mathbb C}^{N_{\rm Tx} \times N_{\rm RF} } $ is a binary matrix indicating the antenna selection in a subarray.

Thus, the complex baseband discrete-time signal at the transmitter can be written as
 \begin{equation}
 {\bf x}[l]= {\bf F}_{\rm RF}   {\bf F}_{{\rm D},l}  {\bf s}_l,
 \label{3}
 \end{equation}
where ${\bf s}_l$  is the normalized symbol sequence corresponding to the $l$-th subpulse with $ {\mathbb E}\{{\bf s}_l {\bf s}_l^H \} = {\bf I}_{N_s} $.

Assuming  $L$ subpulses are contained in one pulse duration, and collecting all $L$ transmit vectors into a matrix ${\bf X} \in {\mathbb C}^{N_{\rm Tx} \times L}$, we have
\begin{equation}
     {\bf X}= {\bf F}_{\rm RF}   [{\bf F}_{{\rm D}, 1}  {\bf s}_1, \cdots, {\bf F}_{{\rm D},L}  {\bf s}_L   ].
\end{equation}

\subsection{Communication Model}
At the recipient vehicle receiver, the   signal corresponding to the $l$-th subpulse is modeled as
\begin{equation}
{\bf c}[l]= {\bf H} {\bf F}_{\rm RF} {\bf F}_{{\rm D},l}  {\bf s}_l + {\bf z}_c[l],
\end{equation}
where $   {\bf H} \in {\mathbb C}^{N_{\rm Rx} \times N_{\rm Tx}}  $    is the  channel state information (CSI)  from
the transmitter to the recipient vehicle and assumed to be known through some channel estimation techniques \cite{li2016optimum, Yin2013A,2004Pilot} such as pilot method.  
{  $ {\bf z}_c[l] $  is additive  Gaussian noise vector with zero mean and  variance  $ \sigma_c^2 $.}   It is assumed that the CSI between the transmitter and the  recipient vehicle 
is modeled as a geometric channel with $ N_{path}  $ paths \cite{sohrabi2016hybrid,2014Low}. Specifically, the channel matrix  $\bf H$ is written as
\begin{equation}
{{\mathbf{H}}  = \sqrt {\frac{1}{N_{path}}} \sum\limits_{l= 1}^{N_{path}}  {\varkappa}_{l}  {{\mathbf{a}}_r}\left( {\phi _{l}^r} \right){{\mathbf{a}}_t^H}{\left( {\phi _{l}^t} \right)}},
\end{equation}
where $ {\varkappa}_{l} \sim {\cal CN}(0,1) $ is the complex factor of the $l$-th path, and the angles of arrival and departure (AoAs/AoDs), $ {\phi _{l}^r}, {\phi _{l}^t}$  are assumed to be uniformly distributed in $ [0, 2\pi )  $. Besides, $ {{\mathbf{a}}_r}\left(  \cdot  \right) $ and $ {{\mathbf{a}}_t}\left( \cdot \right) $ are the  array steering vectors and for uniform linear arrays  (ULAs), they are defined by
\begin{equation} 
   {\bf a}_r(\phi)=\frac{1}{\sqrt{N_{\rm Rx}}}\left[1, e^{\jmath 2\pi d_r \sin \phi/\lambda}, \cdots, e^{\jmath 2\pi d_r (N_{\rm Rx}-1) \sin \phi/\lambda} \right]^T,
   \label{7_1}
   \end{equation}
  \begin{equation} 
{\bf a}_t(\phi)=\frac{1}{\sqrt{N_{\rm Tx}}}\left[1, e^{\jmath 2\pi d_t \sin \phi/\lambda}, \cdots, e^{\jmath 2\pi d_t (N_{\rm Tx}-1) \sin \phi/\lambda} \right]^T,
\end{equation}
where $ \lambda $ represents the carrier wavelength, $d_r$ and $d_t$  denote the antenna spacings at the receiver and transmitter, respectively.

 The  recipient vehicle  adopts an $ N_{\rm Rx}  \times  N_s$ digital combiner ${\bf U}_l =\left[ {{{\bf{u}}_{1,l}}, \cdots ,{{\bf{u}}_{N_s, l}}} \right] $, to  estimate the 
symbol block of the $l$-th subpulse,  then the estimated $ \hat{\bf   s}_{l} $ can be modeled as 
\begin{equation}
\begin{aligned}
\hat{\bf   s}_{l}&= {\bf U}_{l}^H  {\bf c}[l]=  {\bf U}_{l}^H  {\bf H} {\bf F}_{\rm RF} {\bf F}_{{\rm D},l}  {\bf s}_l +  {\bf U}_{l}^H{\bf z}_c[l],
\end{aligned}\label{7}
\end{equation}

For the communication function, we focus on the hybrid precoder design to maximize the  SE, {\color{black}which is used to describe the bandwidth efficiency of communication systems}. Concretely, the SE $R_l({\bf F}_{{\rm D},l}, {\bf F}_{\rm RF}, {\bf U}_l )$ for the $l$-th subpulse is defined as  \cite{sohrabi2016hybrid}
\begin{equation}
\begin{aligned}  R_{l} & ({\bf F}_{{\rm D},l},{\bf F}_{{\rm RF}},{\bf U}_{l})~ {\color{black}[{\rm bits/s/Hz}]} \\
& = \log\Big|{\bf I}_{N_{{\rm Rx}}}+{\bf U}_{l}{\bf C}_{l}^{-1}{\bf U}_{l}^{H}{\bf H}{\bf F}_{{\rm RF}}{\bf F}_{{\rm D},l}{\bf F}_{{\rm D},l}^{H}{\bf F}_{{\rm RF}}^{H}{\bf H}^{H}\Big|, 
\end{aligned}
\label{8}
\end{equation}
 where ${\bf C}_l = {\sigma_c^2}  {\bf U}_{l}^H {\bf U}_{l} $. 

\vspace{-1em}
 \subsection{Radar Model}
 For the radar function,  we assume that the radar receive array with $N_{\rm Rad}$ elements adopts full-digital beamforming structure, and consider a scenario where the radar receiver needs to detect the target vehicle of interest in the presence of clutter. In mmWave band, the scattering of target is extended in distance due to the high range resolution. To be more specific, let $\theta_t$ be the angle of a generic extended target and $t(k), k=0, \cdots, L_{\rm tar}-1$ be  the finite impulse response (FIR) of the extended target with $L_{\rm tar}$ being the  
 support length of the FIR \cite{4840496,4200705}. 
 Then, the received vector is modeled as
 \begin{equation}
 {\bf r}[n  ]={\bf H}_t (\theta_t) e^{\jmath 2 \pi n f_d/f_s   } \sum\limits_{l = 1}^L t(n  -l) {\bf x}[l] + {\bf j}[n ] + {\bf{z}}_r[n ],
 \label{eq:radarmodel}
 \end{equation}
 where ${\bf H}_t (\theta_t) = {\bf a}_{Rr}(\theta_t) {\bf a}_t^H(\theta_t)$ is the spatial steering matrix with $ {\bf a}_{Rr}(\theta_t) $ being the radar receive response  vector similar to \eqref{7_1}, $f_d =\frac{2v_r}{\lambda}$ is the Doppler shifts of the target with $v_r$ being the  radial  velocity of the target, $f_s$ is the sampling frequency,    ${\bf z}_r[n]$ is a zero-mean  Gaussian noise vector  with variance  $ \sigma_{r}^2 $, and ${\bf j}[n]$ is interference term from the stationary clutters. Assuming that the clutter is divided into $K$ clutter bins located at $\theta_i,\forall i=1,\cdots,K$, then ${\bf j}[n]$ is expressed as 
 \begin{equation}
 {\bf j}[n]= \sum\limits_{i = 1}^K  {\bf H}_i (\theta_i) \sum\limits_{l = 1}^L j_i(n-l) {\bf x}[l],
 \end{equation}
 where  ${\bf H}_i (\theta_i) = {\bf a}_{Rr}(\theta_i) {\bf a}_t^H(\theta_i)$ is the spatial steering matrix of the $i$-th clutter bin and $ j_i[k], k=0,\cdots,L_{c, i}-1 $ denotes the FIR of the $i$-th clutter bin with $L_{c,i}$ being the support length. 
 
We define ${\bf t}= [t(0),\cdots,t(L_{\rm tar}-1) ]^T$ and ${\bf j}_i=[j_i(0), \cdots, j_i(L_{c,i}-1)]^T$ and assume that both $\bf t$ and $\{{\bf j}_i\}$ are zero mean random vectors with covariance matrix being ${\bf \Sigma}_t={\mathbb E}\{{\bf t} {\bf t}^H \}$ and ${\bf \Sigma}_{c,i} ={\mathbb E}\{{\bf j}_i {\bf j}_i^H\}$, respectively\footnote{\color{black}Here, we assume the covariance matrices of target and clutter are known. The assumption of the availability of this prior information
can be obtained via a knowledge-based approach \cite{guerci2010cognitive,chen2009mimo}.}. Let $L_{\rm obs}=L+\max\{L_{\rm tar},\{L_{c,i}\}\}-1$ being the receiver observation length.
After defining ${\bf R}=[ {\bf r}[1],\cdots,  {\bf r}[L_{\rm obs}]  ]\in {\mathbb C}^{N_{\rm Rx}  \times L_{\rm obs}}$ and ${\bf Z}_r=[ {\bf z}_{r}[1],\cdots,  {\bf r}_{r}[L_{\rm obs}]  ]\in {\mathbb C}^{N_{\rm Rx}  \times L_{\rm obs}}$, the model can be written in the matrix form as follows:
\begin{equation}
 {\bf R}={\bf H}_t (\theta_t) {\bf X} {\bf T} {  {\cal F}}_d  +\sum\limits_{i = 1}^K {\bf H}_i (\theta_i) {\bf X} {\bf J}_i + {\bf{Z}}_r,
\end{equation}
where
\[{{\bf{T}} = \left[ {\begin{array}{*{20}{c}}
 	{t(0)}& \cdots &{t({L_{{\rm{tar}}}}-1)}&{}&0\\
 	{}& \ddots &{}& \ddots &{}\\
 	0&{}&{t(0)}& \cdots &{t({L_{{\rm{tar}}}}-1)}
 	\end{array}} \right] \in {{\mathbb C}^{L \times {L_{{\rm{obs}}}}}}, }\]
\[{{{\bf{J}}_i} = \left[ {\begin{array}{*{20}{c}}
 	{{j_i}(0)}& \cdots &{{j_i}({L_{c,i}} - 1)}&{}&0\\
 	{}& \ddots &  & \ddots &{}\\
 	0&{}&{{j_i}(0)}& \cdots &{{j_i}({L_{c,i}} - 1)}
 	\end{array}} \right] \in {{\mathbb C}^{L \times {L_{{\rm{obs}}}}}}. }\]
and ${{  {\cal F}}_d ={\rm diag}\{e^{\jmath 2 \pi   f_d/f_s   }, \cdots, e^{\jmath 2 \pi L_{\rm obs} f_d/f_s   } \}}$. 
 
The received signal $  {\bf R}  $  is filtered  via the receive beamformer ${\bf V} \in {\mathbb C}^{N_{\rm Rad}\times L_{\rm obs}}$, then the output SINR\footnote{\color{black}For target detection applications, the detection probability ($P_d$) of the target can be evaluated as $P_d={Q}(\sqrt{2{\rm SINR}}, \sqrt{-2\ln P_{fa}})$\cite{5263002}, where $Q(\cdot, \cdot)$ is the Marcum Q function of order 1 and $P_{fa}$ is the false alarm probability. Thereby, for a specified value $P_{fa}$, the maximization of $P_d$ is equivalent to the maximization of SINR.}  can be written as
\begin{equation}
\begin{aligned}
&{\rm SINR}({\bf F}_{\rm RF}, {\bf F}_{\rm D}, {\bf V}) \\
& = \frac{{{\mathbb E}\left\{ {{{\left| {{\rm{Tr}}\left\{ {{{\bf{V}}^H}{{\bf{H}}_t}({\theta _t}){\bf{XT}}} {  {\cal F}}_d \right\}} \right|}^2}} \right\}}}{{\mathop \sum \limits_{i = 1}^K {\mathbb E} \left\{ {{{\left| {{\rm{Tr}}\left\{ {{{\bf{V}}^H}{{\bf{H}}_i}({\theta _i}){\bf{X}}{{\bf{J}}_i}} \right\}} \right|}^2}} \right\} + {\mathbb E}\left\{ {{{\left| {{\rm{Tr}}\left\{ {{{\bf{V}}^H}{{\bf{Z}}_r}} \right\}} \right|}^2}} \right\}}},
\end{aligned}\label{13}
\end{equation}
where ${\bf F}_{\rm D}=[{\bf F}_{{\rm D},1}, \cdots, {\bf F}_{{\rm D},L}]$. The following proposition will be used when designing the HBF and radar receive filter. 

{\it Proposition 1:  The SINR in \eqref{13} can be equivalently expressed as}
\begin{subequations}
      \begin{align}{\rm SINR}({\bf F}_{{\rm RF}},{\bf F}_{{\rm D}},{\bf V} & )=\dfrac{{\bf v}^{H}{{\bf {\Theta}}_{t}}({{\bf {F}}_{{\rm {RF}}}},{{\bf {F}}_{{\rm {D}}}}){\bf v}}{{\bf v}^{H}{{\bf {\Theta}}_{c}}({{\bf {F}}_{{\rm {RF}}}},{{\bf {F}}_{{\rm {D}}}}){\bf v}+\sigma_{r}^{2}{\bf v}^{H}{\bf v}}\label{14a}\\
 & =\dfrac{\sum\limits _{l=1}^{L}{{\rm {Tr}}\left\{ {{{\bf {F}}_{l}}{\bf {F}}_{l}^{H}{{\bf {\Phi}}_{t}}[l,l]}\right\} }}{\sum\limits _{l=1}^{L}{{\rm {Tr}}\left\{ {{{\bf {F}}_{l}}{\bf {F}}_{l}^{H}{{\bf {\Phi}}_{c}}[l,l]}\right\} }+\sigma_{r}^{2}{\bf v}^{H}{\bf v}},\label{14b}
\end{align}
\label{17_1}%
\end{subequations}
where $  {{\bf{\Theta }}_t}({{\bf{F}}_{{\rm{RF}}}},{{\bf{F}}_{\rm{D}}})   $, $  {{\bf{\Theta }}_c}({{\bf{F}}_{{\rm{RF}}}},{{\bf{F}}_{\rm{D}}})  $, $ {{\bf{\Phi }}_t}[l,l] $, $ {{\bf{\Phi }}_c}[l,l] $ are defined in Appendix A,  ${\bf F}_l={\bf F}_{\rm RF} {\bf F}_{{\rm D},l} $  and $ {\bf v}={\rm vec}({\bf V}) $.    
  \begin{IEEEproof}
  	See  Appendix A.
  \end{IEEEproof}
 Predictably, equation  \eqref{14a} will be useful in optimizing radar receive filter $\bf V$ with the fixed  hybrid beamformer $({\bf F}_{{\rm RF}},{\bf F}_{{\rm D}})$. While the alternative  equation  \eqref{14b} is benefit to designing the  $({\bf F}_{{\rm RF}},{\bf F}_{{\rm D}})$ with a given $\bf V$.

\vspace{-1em}
\subsection{Problem Formulation}
In this paper, we assume that the primary function of the DFRC system is to communicate with the recipient vehicle while providing the detection of the target vehicle as the secondary function.  According to above models, a meaningful criterion of jointly optimizing   the  hybrid digital/analog precoder $ \left( {\bf F}_{\rm D}, {\bf F}_{\rm set} \right)  $, communication combiner ${\{{\bf U}_l\}} $ and radar receive  filter $ \bf V $ is to maximize the communication   SE   while keeping the SINR requirement for radar target. Mathematically, our problem of interest can be formulated as 
    \begin{subequations}\label{16}
         \begin{align}
        &  \mathop {\max}\limits_{ {{\bf F}_{\rm D}, {\bf F}_{\rm set} }, {\{{\bf U}_l\}}, {\bf{V}}   } \;\; \sum\limits_{l = 1}^L R_l({\bf F}_{{\rm D},l}, {\bf F}_{\rm RF}, {\bf U}_l )  \\
   &  {\rm{s}}.{\rm{t}}.\quad   {\rm SINR}({\bf F}_{\rm RF}, {\bf F}_{{\rm D}}, {\bf V}) \ge \gamma,\label{16b}\\
&  ~~~~ \quad   {\bf F}_{\rm set} = {\rm diag}\{{f}_{1}, \cdots,{f}_{N_{\rm Tx}}\},{f}_m= A_{m}e^{\jmath \varphi_{m}}, \forall m   \label{16c}\\
& ~~~~ \quad    0\le A_{m} \le {2}/{\sqrt{N_{\rm Tx}}},~ \varphi_{m} \in [0, 2\pi], \forall m,   \label{16d}\\
   &  \;\;\;\;\quad         {\rm{Tr}}\left( {{{\bf{F}}_{{\rm{RF}}}}{{\bf{F}}_{\rm D}}{\bf{F}}_{\rm D}^H{\bf{F}}_{{\rm{RF}}}^H} \right) \le {\cal E}, \label{16e}
     \end{align}
    \end{subequations}
where  the constraint \eqref{16b} is the SINR requirement for the radar target with $ \gamma $  being the SINR threshold, the constraints \eqref{16c} and \eqref{16c} are feasible conditions for the DPS, and the constraint \eqref{16e} is total energy requirement for the DFRC system with $\cal E$ being the total energy budget. 

Note that this optimization problem involves a nonconvex objective function and nonconvex constraints  \eqref{16b}-\eqref{16e}, and hence, it is NP-hard \cite{boyd2004convex} and  challenging to solve.

\section{Hybrid Beamforming Design With  Double Phase shifters Architecture}
In this section, we will solve the HBF design problem in the alternating optimization manner. Concretely, the radar receiver $\bf V$ is optimized for a given $({{\bf F}_{\rm D}, {\bf F}_{\rm set} }, {\{{\bf U}_l\}})$, and in turn, $ ({{\bf F}_{\rm D}, {\bf F}_{\rm set} }, {\{{\bf U}_l\}}) $ are jointly optimized for a given $\bf V$. 
Since the subproblem with respect to $  ({{\bf F}_{\rm D}, {\bf F}_{\rm set} }, {\{{\bf U}_l\}}) $ has a {\it consensus} form, we propose an efficient algorithm by utilizing the consensus-ADMM \cite{boyd2011distributed}. 
  
\subsection{Optimization of radar receiver}
Note that the objective function in \eqref{16} is independent to $\bf V$, and thus, we only need to find a feasible solution $ \bf V $ to meet the SINR requirement \eqref{16b}. To this end,  the radar filter $\bf V$ can be determined by maximizing the SINR value as 
  \begin{equation}
 \max\limits_{ \bf V} ~{\rm SINR}({\bf F}_{\rm RF}, {\bf F}_{D,l}, {\bf V})  =     \dfrac{{\bf v}^H{{\bf{\Theta }}_t}({{\bf{F}}_{{\rm{RF}}}},{{\bf{F}}_{\rm{D}}}) {\bf v}}{{\bf v}^H{{\bf{\Theta }}_{c}}({{\bf{F}}_{{\rm{RF}}}},{{\bf{F}}_{\rm{D}}}) {\bf v}+ \sigma_r^2 {\bf v}^H{\bf v}},   
  \end{equation}
   of which the  optimal solution  ${\bf v}^\star$ can be achieved by taking the generalized eigenvalue decomposition (EVD) of $ \left(  {\bf{\Theta }}_t({{\bf{F}}_{{\rm{RF}}}},{{\bf{F}}_{\rm{D}}}) ,   {\bf{\Theta }}_c({{\bf{F}}_{{\rm{RF}}}},{{\bf{F}}_{\rm{D}}})+ \sigma_r^2 {\bf I}_{{N}_{\rm Rad}N_s } \right)  $, i.e., 
  \begin{equation}\label{16a}
  {\bf v}^\star= {\cal P}\left( {\bf{\Theta }}_t^{-1}({{\bf{F}}_{{\rm{RF}}}},{{\bf{F}}_{\rm{D}}}) \left(  {\bf{\Theta }}_c({{\bf{F}}_{{\rm{RF}}}},{{\bf{F}}_{\rm{D}}})+ \sigma_r^2 {\bf I}_{{N}_{\rm Rad}N_s } \right) \right),
  \end{equation}
  where the operator ${\cal P}(\cdot)$ denotes the principal eigenvector.
  
  \vspace{-1em}
   \subsection{Optimization of hybrid beamformer and combiner}
 For a fixed  $\bf V$, the subproblem with respect to $({{\bf F}_{\rm D}, {\bf F}_{\rm set} }, {\{{\bf U}_l\}})$ is 
  \begin{equation} 
 	\begin{aligned}
 &\mathop {\max}\limits_{ {{\bf F}_{\rm D}, {\bf F}_{\rm set} }, {\{{\bf U}_l\}}   } \;\; \sum\limits_{l = 1}^L R_l({\bf F}_{{\rm D},l}, {\bf F}_{\rm RF}, {\bf U}_l )  \\
 	&  {\rm{s}}.{\rm{t}}.\quad   \eqref{16b},~ \eqref{16c}, ~ \eqref{16d},~ {\rm and}~ \eqref{16e} 
 	\end{aligned}\label{31}
 \end{equation}
   Since $ {\bf F}_{\rm set} $ and  $ {\bf F}_{\rm D} $ are coupled in constraints \eqref{16b} and \eqref{16e}, this subproblem is difficult to solve. By introducing auxiliary variables ${\bf X}_l, {\bf Z}_l\in {\mathbb C}^{N_{\rm Tx} \times {N_s}}, \forall l$, we decouple ${\bf F}_{\rm set}$ and ${\bf F}_{\rm D}$ and recast problem \eqref{31} into 
   \begin{subequations} 
  \begin{align}
  & \mathop {\max}\limits_{ {{\bf F}_{\rm D}, {\bf F}_{\rm set} }, {\{{\bf U}_l\}}, {\{{\bf X}_l\}},{\{{\bf Z}_l\}}   } \;\; \sum\limits_{l = 1}^L R_l({\bf X}_l, {\bf U}_l ), \label{32a}  \\
  &  {\rm{s}}.{\rm{t}}.\quad    \dfrac{\sum\limits_{l = 1}^L {{\rm{Tr}}\left\{ {{{\bf{Z}}_l}{\bf{Z}}_l^H{{\bf{\Phi }}_t}[l,l]} \right\}} }{ \sum\limits_{l = 1}^L {{\rm{Tr}}\left\{ {{{\bf{Z}}_l}{\bf{Z}}_l^H{{\bf{\Phi }}_{c}}[l,l]} \right\}} + \sigma_r^2 {\bf v}^H   {\bf v}   }\ge \gamma, \label{32b}\\
  & \quad ~\quad {\bf X}_l ={\bf Z}_l={\bf F}_{\rm set} {\bf P} {\bf F}_{{\rm D},l},\forall l \label{32c}\\
 & \quad ~\quad    \sum\limits_{l = 1}^L  {\rm{Tr}}\left( {\bf X}_l {\bf{X}}_l^H \right) \le {\cal E},\label{32e} \\
 & \quad ~\quad  ~\eqref{16c},  \eqref{16d},  \label{32d} 
  \end{align}
  \label{32}%
  \end{subequations}
where $R_l({\bf X}_l, {\bf U}_l ) = 
 \log\Big| {\bf I}_{N_{\rm Rx}} + {\bf U}_{l} {\bf C}^{-1}_l {\bf U}_{l}^H {\bf H} {\bf X}_l {\bf X}_l^H  {\bf H}^H  \Big|  $.
 
{\color{black}It is observed that the introduction of auxiliary variables ${\bf X}_l, {\bf Z}_l, \forall l$ results
in decoupling ${\bf F}_{\rm set}$, ${\bf P}  $ and $ {\bf F}_{{\rm D},l}$ in original objective function and    imposing  constraints \eqref{32b} and \eqref{32e} on ${\bf Z}_l, \forall l$ and ${\bf X}_l, \forall l$ respectively. This will enable us to construct the ADMM subproblems with respect to  problem  \eqref{32}, each of which can be solved with a closed form solution.} Concretely, 
placing the   equality constraints $ {\bf X}_l ={\bf Z}_l={\bf F}_{\rm set} {\bf P} {\bf F}_{{\rm D},l} $ into the augmented Lagrangian function of \eqref{32} yields 
      \begin{equation}
      \begin{aligned}
       {\cal L}= \sum_{l=1}^{L} \tilde{\cal L}_l ({\bf W}_l, {\bf X}_l, {\bf U}_l, {\bf Z}_l, {\bf F}_{\rm set},{\bf F}_{{\rm D},l}, {\bf D}_{1,l}, {\bf D}_{2,l}),
      \end{aligned}
  \end{equation}
where $\tilde{\cal L}_i$ is defined as 
\begin{equation}
\begin{aligned} 
& \tilde{{\cal L}}_{l}({\bf W}_{l},{\bf X}_{l},{\bf U}_{l},{\bf Z}_{l},{\bf F}_{{\rm set}},{\bf F}_{{\rm D},l},{\bf D}_{1,l},{\bf D}_{2,l})
= R_l({\bf X}_l, {\bf U}_l ) \\ 
& +\Re\left({\rm Tr}\left\lbrace {\bf D}_{1,l}^{H}\left({\bf X}_{l}-{\bf F}_{{\rm set}}{\bf P}{\bf F}_{{\rm D},l}\right)\right\rbrace \right) +\frac{\rho_{1}}{2} \left\Vert {\bf X}_{l}-{\bf F}_{{\rm set}}{\bf P}{\bf F}_{{\rm D},l}\right\Vert _F^{2} \\
& +\Re\left({\rm Tr}\left\lbrace {\bf D}_{2,l}^{H}\left({\bf X}_{l}-{\bf Z}_{l}\right)\right\rbrace \right)+\frac{\rho_{2}}{2} \left\Vert {\bf X}_{l}-{\bf Z}_{l}\right\Vert _F^{2},
\end{aligned}
\end{equation}
where $ {\bf D}_{1,l}, {\bf D}_{2,l} \in {\mathbb C}^{N_{\rm Tx} \times N_s} $ are dual variables corresponding to the equalities $  {\bf X}_l = {\bf F}_{\rm set} {\bf P} {\bf F}_{{\rm D},l}  $ and $ {\bf X}_l ={\bf Z}_l $, respectively, and ${\rho_{1}}, \rho_{2}>0$ are the penalty parameters.

To fulfill the convergence requirements of the consensus-
ADMM, we split the optimized primal variables into two blocks $\left( {\bf W}_l, {\bf X}_l, {\bf U}_l    \right)  $ and $ \left( {\bf Z}_l, {\bf F}_{\rm set},{\bf F}_{{\rm D},l}    \right)  $. In what follows, we  shall present the update procedures of  the two primal blocks and dual block  $ \left( {\bf D}_{1,l}, {\bf D}_{2,l}\right)$. 
\subsubsection{Optimization of $  \left( {\bf W}_l, {\bf X}_l, {\bf U}_l    \right)  $}
For fixed $ \left( {\bf Z}_l, {\bf F}_{\rm set},{\bf F}_{{\rm D},l}    \right)  $ and $ \left( {\bf D}_{1,l}, {\bf D}_{2,l}    \right)$, $ \left( {\bf W}_l, {\bf X}_l, {\bf U}_l    \right) $ is updated by solving
\begin{equation} 
  \begin{aligned} & \mathop{\min}\limits _{\{{\bf W}_{l}\},\{{\bf X}_{l}\},\{{\bf U}_{l}\}} &  & \sum_{l=1}^{L}\tilde{{\cal L}}_{l}({\bf W}_{l},{\bf X}_{l},{\bf U}_{l},{\bf Z}_{l},{\bf F}_{{\rm set}},{\bf F}_{{\rm D},l},{\bf D}_{1,l},{\bf D}_{2,l})\\
 & \quad\quad\;{\text{s.t.}}. &  & \sum\limits _{l=1}^{L}{\rm {Tr}}\left({\bf X}_{l}{\bf {X}}_{l}^{H}\right)\le {\cal E}.
\end{aligned}
  \label{37}
  \end{equation}
  
Nevertheless, 
It is still difficult to find the solution of \eqref{37} due to  the nonconvex function $R_l({\bf X}_l, {\bf U}_l )$.
To solve problem \eqref{37}, the following theorem is useful. 

    {\color{black}
\textit{Theorem 1}:
Based on the WMMSE method\cite{shi2011iteratively},
maximizing $  \sum\limits_{l = 1}^L R_l({\bf X}_l, {\bf U}_l )$ can be equivalently replaced by,  
  \begin{equation}
\min f({\bf W}_l, {\bf X}_l, {\bf U}_l )=   \sum\limits_{l = 1}^L  {\rm Tr}\left\lbrace {\bf E}_l ({\bf X}_l, {\bf U}_l )  {\bf W}_l \right\rbrace - \log\Big|  {\bf W}_l \Big|,
  \end{equation}
  where  $ {\bf W}_l $ is the weight matrix, and $ {\bf E}_l({\bf X}_l, {\bf U}_l )$  is the MSE matrix, given by
    \begin{equation}
\begin{aligned} 
& {\bf E}_{l}({\bf X}_{l},{\bf U}_{l}) \\
& = \left({\bf I}_{N_{s}}-{\bf U}_{l}^{H}{\bf H}{\bf X}_{l}\right)\left({\bf I}_{N_{s}}-{\bf U}_{l}^{H}{\bf H}{\bf X}_{l}\right)^{H}+\sigma_{c}^{2}{\bf U}_{l}^{H}{\bf U}_{l}.
\end{aligned}
\label{34}
\end{equation}}
\begin{IEEEproof}
    {\color{black} See Appendix B.}  
\end{IEEEproof}

By doing so, a coordinate descent (CD)-type algorithm is utilized to update the variables iteratively. 
Specifically, the update of ${\bf U}_l$ is obtained by solving 
\begin{equation} 
\begin{aligned}
&  \mathop {\min}\limits_{    {\bf U}_l  } \;\; {\rm Tr}\left\lbrace {\bf E}_l ({\bf X}_l, {\bf U}_l )  {\bf W}_l \right\rbrace.
\end{aligned}\label{38}
\end{equation}
According to \eqref{34},  its optimal solution of ${\bf U}_l$ can be attained via the first-order optimality condition given by 
      \begin{equation} 
  \begin{aligned}
 {\bf U}_l=\left(     {\bf H}   {\bf X}_l {\bf X}_l^H    {\bf H}^H +   \sigma_c^2 {\bf I}_{\rm Rx}   \right)^{-1}  {\bf H}   {\bf X}_l.
  \end{aligned}\label{39}
  \end{equation}
  
  The update of ${\bf W}_l$ is obtained by solving 
        \begin{equation} 
  \begin{aligned}
  \mathop {\min}\limits_{    {\bf W}_l  } \;\; {\rm Tr}\left\lbrace {\bf E}_l ({\bf X}_l, {\bf U}_l )  {\bf W}_l \right\rbrace -  \log  \Big|  {\bf W}_l \Big|,
  \end{aligned}\label{40}
  \end{equation}
  which has the optimal solution given by
\begin{equation} 
  \begin{aligned}
  {\bf W}_l= {\bf E}_l^{-1}({\bf X}_l, {\bf U}_l ) = \left({\bf I}_{N_s}- {\bf X}_l^H {\bf H}^H   {\bf U}_l \right)^{-1}.
  \end{aligned}\label{41}
  \end{equation}

To proceed, the update of ${\bf X}_l$ is obtained by solving 
        \begin{equation} 
\begin{aligned}
&\mathop {\min}\limits_{    {\bf X}_l  }  {\rm Tr}\left\lbrace {\bf E}_l ({\bf X}_l, {\bf U}_l )  {\bf W}_l \right\rbrace  + \Re\left({\rm Tr}\left\lbrace{\bf D}_{1,l}^H \left(  {\bf X}_l-  {\bf F}_{\rm set} {\bf P} {\bf F}_{{\rm D},l} \right)  \right\rbrace    \right)\\
  &\qquad +\frac{\rho_{1}}{2} \left\|    {\bf X}_l-  {\bf F}_{\rm set} {\bf P} {\bf F}_{{\rm D},l} \right\|_F^2+ \Re\left({\rm Tr}\left\lbrace{\bf D}_{2,l}^H \left(  {\bf X}_l-  {\bf Z}_{l}  \right)     \right\rbrace \right) \\
  & \qquad +\frac{\rho_{2}}{2} \left\|    {\bf X}_l- {\bf Z}_l\right\|_F^2\\
    &  {\rm{s}}.{\rm{t}}.\quad     \sum\limits_{l = 1}^L  {\rm{Tr}}\left( {\bf X}_l {\bf{X}}_l^H \right) \le {\cal E}.
\end{aligned}\label{42}
\end{equation}
The following theorem provides the solution to problem \eqref{42}. 

{\textit{Theorem 2}: The optimal solution to problem \eqref{42} can be found via the Karush-Kuhn-Tucker (KKT) conditions.}
  \begin{IEEEproof}
  	See  Appendix C.
  \end{IEEEproof}

    \subsubsection{Optimization of  $ \left( {\bf Z}_l, {\bf F}_{\rm set},{\bf F}_{{\rm D},l}    \right)  $}
  For fixed $ \left( {\bf W}_l, {\bf X}_l, {\bf U}_l    \right)  $ and $ \left( {\bf D}_{1,l}, {\bf D}_{2,l}    \right)$,  $ \left( {\bf Z}_l, {\bf F}_{\rm set},{\bf F}_{{\rm D},l}    \right) $  is updated by solving
   \begin{equation} 
  \begin{array}{cl}
\mathop{\min}\limits _{{\bf Z}_{l},{\bf F}_{{\rm set}},{\bf F}_{{\rm D},l}} & \sum_{l=1}^{L}\tilde{{\cal L}}_{l}\left({\bf W}_{l},{\bf X}_{l},{\bf U}_{l},{\bf Z}_{l},{\bf F}_{{\rm set}},{\bf F}_{{\rm D},l},{\bf D}_{1,l},{\bf D}_{2,l}\right)\\
\text{s. t. } & \sum\limits _{l=1}^{L}{\rm Tr}\left({\bf Z}_{l}{\bf Z}_{l}^{H}{\bf M}[l,l]\right)\ge\alpha, \eqref{16c},\eqref{16d},
\end{array}\label{49}
  \end{equation}
where $ {\bf M}[l,l]={{\bf{\Phi }}_t}[l,l] -\gamma {{\bf{\Phi }}_c}[l,l] $, and $\alpha=\gamma \sigma_r^2 \|{\bf v}\|^2$ . We note  that the CD method is able to solve the problem \eqref{49}. Specifically, the update of ${\bf Z}_l$ needs solving   
     \begin{equation} 
  \begin{aligned}
  &  \mathop {\min}\limits_{  {\bf Z}_l  } \;\; \sum_{l=1}^{L}  \Re\left({\rm Tr}\left\lbrace {\bf D}_{2,l}^H \left(  {\bf X}_l-  {\bf Z}_{l}  \right) \right\rbrace     \right) +\frac{\rho_{2}}{2} \left\|    {\bf X}_l- {\bf Z}_l\right\|_F^2\\
  &  {\rm{s}}.{\rm{t}}.\quad     \sum\limits_{l = 1}^L  {\rm{Tr}}\left( {\bf Z}_l {\bf{Z}}_l^H  {\bf M}[l,l]\right) \ge \alpha.
  \end{aligned}\label{50}
  \end{equation}
  
Similar to the solution to problem \eqref{42}, the following theorem is useful to give the solution to problem \eqref{50}.
  
{\textit{Theorem 3}: The optimal solution to problem \eqref{50} is obtain by analyzing the KKT conditions.}
\begin{IEEEproof}
  See  Appendix D.
\end{IEEEproof}

The  variables ${\bf F}_{{\rm D},l}, \forall l$ are updated in parallel by solving
\begin{equation} 
\begin{aligned} \mathop{\min}\limits _{{\bf F}_{{\rm D},l}} & \quad  \sum_{l=1}^{L}\Re\left({\rm Tr}\left\lbrace {\bf D}_{1,l}^{H}\left({\bf X}_{l}-{\bf F}_{{\rm set}}{\bf P}{\bf F}_{{\rm D},l}\right)\right\rbrace \right) \\
& \quad +\frac{\rho_{1}}{2}\left\Vert {\bf X}_{l}-{\bf F}_{{\rm set}}{\bf P}{\bf F}_{{\rm D},l}\right\Vert ^{2},
\end{aligned}
\label{51}
\end{equation}
whose closed-form solution is 
\begin{equation} 
   \begin{aligned}
 {\bf F}_{{\rm D},l}=&\left(   {\bf P}^H  {\bf F}_{\rm set}^H {\bf F}_{\rm set}{\bf P}  \right)^{-1}  {\bf P}^H  {\bf F}_{\rm set}^H  \left(  \dfrac{1}{\rho_1}{\bf D}_{1,l} +{\bf X}_{1,l}   \right)  \\
 = & {\rm diag}^{-1}\Big(\sum\limits_{i = 1}^{{N_{{\rm{Tx}}}}} {A_i^2p_{i,1}}, \cdots,  \sum\limits_{i = 1}^{{N_{{\rm{Tx}}}}} {A_i^2p_{i,{N_{{\rm{RF}}}}}} \Big)  \\
 & \qquad \times {\bf P}^H  {\bf F}_{\rm set}^H  \left(  \dfrac{1}{\rho_1}{\bf D}_{1,l} +{\bf X}_{1,l}   \right).
\end{aligned}
\label{44_1}
\end{equation} 
    
The variable ${\bf F}_{{\rm set}}$ is updated by the following problem:  
\begin{equation} 
\begin{array}{cl}
    \mathop{\min}\limits _{{\bf F}_{{\rm set}}} & \sum_{l=1}^{L}\left\Vert {\bf X}_{l}-{\bf F}_{{\rm set}}{\bf P}{\bf F}_{{\rm D},l}+\frac{1}{\rho_{1}}{\bf D}_{1,l}\right\Vert ^{2}\\
    \text{s.t.} & {\bf F}_{{\rm set}}={\rm diag}\left(f_{1},\cdots,f_{N_{{\rm Tx}}}\right),f_{m}=A_{m}e^{\jmath\varphi_{m}},\forall m\\
    & 0\le A_{m} \le {2}/{\sqrt{N_{\rm Tx}}},\varphi_{m}\in[0,2\pi],\forall m.
\end{array}
\label{59}
\end{equation}
Let ${\bf \Pi}_l =   {\bf X}_l +\frac{1}{\rho_1}  {\bf D}_{1,l} $ and $ {\bf Y}_l= {\bf P} {\bf F}_{{\rm D},l}  $, problem \eqref{59} can be decomposed  into
\begin{equation} 
    \begin{aligned}
     \mathop {\min}\limits_{ \{f_m\}} & \sum_{l=1}^{L}  \left\|    {\bf \Pi}_l[m,:] -  f_m  {\bf Y}_l[m,:]  \right\|^2_2 \\
     {\rm{s}}.{\rm{t}}.  ~  &   {   f}_m= A_{m}e^{\jmath \varphi_{m}},  ~     0\le A_{m} \le {2}/{\sqrt{N_{\rm Tx}}}, \forall m,
    \end{aligned}\label{60}
    \end{equation}
whose closed-form solution is 
  \begin{equation}
  	{A_m} = \left\{ \begin{array}{ll}
  	\frac{{\left| {\sum\limits_{l = 1}^L {{{\bf{\Pi }}_l}[m,:]{\bf{Y}}_l^H[m,:]} } \right|}}{{\sum\limits_{l = 1}^L {\left\| {{{\bf{Y}}_l}[m,:]} \right\|_2^2} }} ,& \frac{{\left| {\sum\limits_{l = 1}^L {{{\bf{\Pi }}_l}[m,:]{\bf{Y}}_l^H[m,:]} } \right|}}{{\sum\limits_{l = 1}^L {\left\| {{{\bf{Y}}_l}[m,:]} \right\|_2^2} }} \le \frac{2}{{\sqrt {{N_{{\rm{Tx}}}}} }} \\
  	\frac{2}{{\sqrt {{N_{{\rm{Tx}}}}} }}, & \frac{{\left| {\sum\limits_{l = 1}^L {{{\bf{\Pi }}_l}[m,:]{\bf{Y}}_l^H[m,:]} } \right|}}{{\sum\limits_{l = 1}^L {\left\| {{{\bf{Y}}_l}[m,:]} \right\|_2^2} }} > \frac{2}{{\sqrt {{N_{{\rm{Tx}}}}} }}\;\;
  	\end{array} \right.\label{47}
  \end{equation}
  and 
    \begin{equation}
  {\varphi_{m}} =  \angle \left(  {\sum\limits_{l = 1}^L {{{\bf{\Pi }}_l}[m,:]{\bf{Y}}_l^H[m,:]} }  \right). 
  \end{equation}
  After obtaining $ 	{A_m}  $ and $   {\varphi_{m}}  $, the phase values
  of phase shifters \#1 and \#2 in the DPS element are 
  \begin{subequations}
  	\begin{align}
  	 {\psi _{1,m}} &= {\varphi _m} + \arccos \left( {{A_m}/2} \right),\\
  	 {\psi _{2,m}} &= {\varphi _m} - \arccos \left( {{A_m}/2} \right).
  	\end{align}
  \end{subequations}

    \subsubsection{Optimization of  $ \left( {\bf D}_{1,l},  {\bf D}_{2,l}   \right)  $}
For fixed $ \left( {\bf W}_l, {\bf X}_l, {\bf U}_l    \right)  $ and $  \left( {\bf Z}_l, {\bf F}_{\rm set},{\bf F}_{{\rm D},l}    \right) $, $ \left( {\bf D}_{1,l},  {\bf D}_{2,l}   \right)  $  are updated by  \cite{boyd2011distributed}:
       \begin{equation} 
\begin{aligned}
 {\bf D}_{1,l}&= {\bf D}_{1,l} + \rho_1 \left(  {\bf X}_l -  {\bf F}_{\rm set} {\bf P}{\bf F}_{{\rm D},l} \right), \\
  {\bf D}_{2,l}&= {\bf D}_{2,l} + \rho_1 \left(  {\bf X}_l -   {\bf Z}_{l} \right). 
\end{aligned}\label{aa}
\end{equation}

The consensus-ADMM for  solving problem \eqref{31} is sumerized in Algorithm \ref{alg:alg4_1}. 
  \begin{algorithm}
  	\caption{Consensus-ADMM for solving problem \eqref{31}}
  	\label{alg:alg4_1}
  	\begin{algorithmic}[1]
  		\STATE \textbf{Input:}Initial variables ${\bf F}_{\rm set}(0), {{\bf{F}}_{D,l}}(0)$, ${{\bf{X}}_{l}}(0)$, ${{\bf{Z}}_{l}}(0)$,  ${{\bf{D}}_{1,l}}(0)$,  ${{\bf{D}}_{2,l}}(0)$ and $ {\rho}_{1}, \rho_2 >0 $
  		\STATE Set $ k=0$
  		\REPEAT
  		\STATE  Update $ {\bf U}_l(k+1)$ according to \eqref{39}
  		\STATE  Update ${\bf W}_l(k+1) $ in parallel by \eqref{41}
  		\STATE  Compute $\mu^{\rm opt}$ via the bisection method, and update $ {\bf X}_{l} (k+1), \forall l $ in parallel by \eqref{46}
  		\STATE   Compute $\nu^{\rm opt}$ via the Newton method, and update $ {\bf Z}_{l} (k+1), \forall l $ in parallel by \eqref{53}
  			\STATE Update ${\bf F}_{{\rm D},l}(k+1), \forall l$ according to \eqref{44_1}
  		\STATE Update ${\bf F}_{{\rm set}}(k+1) $ according to \eqref{47}
  		\STATE Update ${\bf D}_{{\rm 1},l}(k+1) $ and ${\bf D}_{{\rm 2},l}(k+1) $  according to \eqref{aa}
  		\STATE  $ k=k+1 $
  		\UNTIL    $ k=N_{\rm ADMM}^{\max}$.
  		\STATE  {\bf Output:} ${\bf F}_{\rm set}^\star= {\bf F}_{\rm set}(k), {{\bf{F}}_{D,l}^\star} ={\bf{F}}_{D,l}(k)$
  	\end{algorithmic}
  \end{algorithm}

Finally, the proposed THEREON algorithm, which jointly optimize radar reciever and hybrid beamformer,   is summarized in Algorithm \ref{alg:alg4_2}.
  \begin{algorithm}
  	\caption{join{T} {H}ybrid b{E}amforming and {R}adar r{E}ceiver {O}ptimizatio{N} (THEREON)}
  	\label{alg:alg4_2}
  	\begin{algorithmic}[1]
  		\STATE \textbf{Input:}Initial variables   ${\bf F}_{\rm set}(0), {{\bf{F}}_{D,l}}(0)$ and iteration number $N_{\rm THER}^{\max}$
  		\STATE Set $ t=0$
  		\REPEAT
  		\STATE  Update $ {\bf V}(t+1)$ according to \eqref{16a}
  		\STATE  Update ${\bf F}_{\rm set}(t+1), {{\bf{F}}_{D,l}}(t+1)$ by using Algorithm \ref{alg:alg4_1}
  		\STATE  $ t=t+1 $
  		\UNTIL    $ t=N_{\rm THER}^{\max}$
  		\STATE  {\bf Output:} ${\bf V}^\star= {\bf V}(t),  {\bf F}_{\rm set}^\star= {\bf F}_{\rm set}(t), {{\bf{F}}_{D,l}^\star} ={\bf{F}}_{D,l}(t)$  
  	\end{algorithmic}
  \end{algorithm}
    	 \vspace{-1em}
    
 \vspace{-1em}
  \subsection{Complexity Analysis}
  We  first analyze  the  computational complexity of the consensus-ADMM method for updating the hybrid beamformer ${\bf F}_{\rm set} $ and $ {{\bf{F}}_{\rm D }} $.  Note that in each iteration of the proposed  consensus-ADMM,  the main computational complexity   is caused  by updating six variables, i.e. ${\bf U}_l$,   ${\bf W}_l$, $  {\bf X}_l $, $  {\bf Z}_l $,  $ {\bf F}_{\rm set} $ and ${\bf F}_{{\rm D},l} $. Updating  ${\bf U}_l$ and  ${\bf W}_l$ based on  \eqref{39} and \eqref{41}   need   complexities of $ {\cal O}\left(  N_s^2N_{\rm Tx} +N_{\rm Rx}^3 \right)  $ and $ {\cal O}\left(  N_s^2N_{\rm Tx} +N_s^3 \right)  $, respectively.   Updating  $  {\bf X}_l $ needs computing  $ {\bf \Lambda}  $ using the bisection method     with  complexities of $ {\cal O}\left(  N_{\rm Tx}L \log_2(n) \right)  $. Updating $  {\bf Z}_l $ needs computing  $ {\bf \nu}  $ using the Newton method     with  complexities of $ {\cal O}\left(  N_{\rm Tx}L \log_2(n) \right)  $, updating  ${\bf F}_{{\rm D},l} $ based on \eqref{44_1}  needs  a complexity of $ {\cal O}\left(N_{\rm Tx}\right)  $ and updating  ${\bf F}_{\rm set}$  based on \eqref{47} needs a complexity of $ {\cal O}\left(    L N_s N_{\rm Tx}    \right)   $.  To  summarize, the overall complexity of  the consensus-ADMM is   $  {\cal O} \Big(N_{\rm ADMM}^{\max}( N_s^2N_{\rm Tx}   + N_{\rm Rx}^3  +N_{\rm Tx}L \log_2(n)+  L N_s N_{\rm Tx}  ) \Big) $. 
 While the complexity of the update of radar recceiver $\bf v$ is $  {\cal O} \Big(  N_s^3 N_{\rm Rad}^3    \Big) $. Overall, the complexity of the THEREON algorithm is $  {\cal O} \Bigg(N_{\rm THER}^{\max} \Big( N_s^3 N_{\rm Rad}^3+ N_{\rm ADMM}^{\max}( N_s^2N_{\rm Tx}   + N_{\rm Rx}^3  +N_{\rm Tx}L \log_2(n)+  L N_s N_{\rm Tx}  ) \Big) \Bigg) $.

\section{Extension to Hybrid Beamforming Design for MU-MISO}
In this section, we extend the proposed method in Sec. III to the hybrid beamforming design for a MU-MISO system in which a transmitter with $N_{\rm Tx}$ antennas and $ N_{\rm RF} $ RF chains serves $N_U$ non-cooperative single-antenna users.
  
In such a system, the transmitted signal at the $l$-th subpulse is given by 
   \begin{equation}
  {\bf x}[l]= {\bf F}_{\rm RF}   {\bf F}_{{\rm D},l}  {\bf s}_l,
  \label{50_1}
  \end{equation}
where ${\bf s}_l = [s_l[1], \cdots, s_l[N_U]]^T$ with $ s_l[u] $ being the intended data symbol for user $ u $   at the subpulse $l$.  with $ {\mathbb E}\{{\bf s}_l {\bf s}_l^H \} = {\bf I}_{N_U} $.
The received signal of the user $n$ at the $l$-th subpulse is 
   \begin{equation}
   { c}_n[l]= {\bf h}_n^H {\bf F}_{\rm RF} {\bf F}_{{\rm D},l}  {\bf s}_l + { z}_n[l].
   \end{equation}
  where  ${ z}_n[l]  $ is the additive white Gaussian noise with variance of $\sigma_n^2$.
  
  The SE for user $u$ in $l$-th subpulse is defined as 
  \begin{equation}
  \begin{aligned}
  {R_l}[n] = \log \left( {1 + \frac{{{{\left| {{\bf{h}}_n^H{{\bf{F}}_{{\rm{RF}}}}{{\bf{F}}_{{\rm{D}},l}}[n]} \right|}^2}}}{{\sigma _n^2 + \sum\nolimits_{i \ne n} {{{\left| {{\bf{h}}_n^H{{\bf{F}}_{{\rm{RF}}}}{{\bf{F}}_{{\rm{D}},l}}[i]} \right|}^2}} }}} \right),
  \end{aligned}
  \end{equation}
  where $ {{{\bf{F}}_{{\rm{D}},l}}[n]} $ is the $n$-th column of the matrix ${\bf F}_{{\rm D},l} $.
  Thus, the hybrid beamforming design problem is formulated as 
      \begin{subequations}\label{53_1}
  	\begin{align}
  	&  \mathop {\max}\limits_{ {{\bf F}_{\rm D}, {\bf F}_{\rm set} }, {\bf{V}}   } \;\; \sum\limits_{l = 1}^L \sum\limits_{n = 1}^{N_U}   \beta_n  {R_l}[n]   \\
  	&  {\rm{s}}.{\rm{t}}.\quad   {\rm SINR}({\bf F}_{\rm RF}, {\bf F}_{\rm D}, {\bf V}) \ge \gamma,\label{53b}\\
  	&  ~~~~ \quad   {\bf F}_{\rm set} = {\rm diag}\{ {    f}_{1}, \cdots,    {  f}_{N_{\rm Tx}  } \}, {   f}_m= A_{m}e^{\jmath \varphi_{m}}, \forall m   \label{53c}\\
  	& ~~~~ \quad    A_{m}\in [0, {2}/{\sqrt{N_{\rm Tx}}}],~ \varphi_{m} \in [0, 2\pi], \forall m,   \label{53d}\\
  	&  \;\;\;\;\quad         {\rm{Tr}}\left( {{{\bf{F}}_{{\rm{RF}}}}{{\bf{F}}_{\rm D}}{\bf{F}}_{\rm D}^H{\bf{F}}_{{\rm{RF}}}^H} \right)\le {\cal E}, \label{53e}
  	\end{align}
  \end{subequations}
where the parameter $\beta_n$ represents the priority of the user $n$. 
  
Relative to   \eqref{16}, the problem of hybrid beamforming design for MU-MISO system has a major difference: For the MU-MISO scenario, there is multi-user interference (MUI) term in the  spectral efficiency expression.  
  
  We note that the difference  between problem \eqref{53_1} and \eqref{16} lies in the objective function while the constraint set is unchanged. Since the consensus-ADMM was used to avoid the coupling in the constraints, it can be used herein as well.
Here we briefly describe the corresponding solving procedure. We first introduce ${\bf X}_l ={\bf Z}_l={\bf F}_{\rm set} {\bf P} {\bf F}_{{\rm D},l}, \forall l$ to decouple $ {\bf F}_{\rm set} $ and  $ {\bf F}_{\rm D} $ in problem \eqref{53_1}. Based on the WMMSE framework, the objective function in \eqref{53_1} can be expressed as,
    \begin{equation}
  f( {  w}_{l,n}, {\bf X}_l, { u}_{l,n} )=   \sum\limits_{l = 1}^L  \sum\limits_{n = 1}^{N_U}  \beta_n \left(  {  w}_{l,n}  {  e}_{l,n} - \log({  w}_{l,n} )  \right),
  \end{equation}
where $ {w}_{l,n} $ is the positive  weight for user $n$ at the subpulse $l$, $ {  e}_{l,n}$  is the MSE error, given by  
  \begin{equation}
{  e}_{l,n}= {\left| {{u_{l,n}}{\bf{h}}_n^H{{\bf{X}}_l}[n] - 1} \right|^2} + {\sum\nolimits_{i \ne n} {{{\left| {{u_{l,n}}{\bf{h}}_n^H{{\bf{X}}_l}[i]} \right|}^2}}}  + \sigma _n^2{\left| {{u_{l,n}}} \right|^2}
  \label{55_1}
  \end{equation}
Similar to the solution procedure demonstrated in Section III, we provide the update solutions of $ {  u}_{l,n} $ $ {  w}_{l,n} $ and ${\bf X}_l$ directly omitting the derivation details. 

1) Calculate the receiver combining filter as 
  \begin{equation}
  {u_{l,n}} = \frac{{  {{\bf{X}}_l^H}[n] {\bf{h}}_n  }}{{\sum\limits_{i = 1}^{{N_U}} {{{\left| {{\bf{h}}_n^H{{\bf{X}}_l}[i]} \right|}^2} + \sigma _n^2} }}.
  \label{56_1}
  \end{equation}
  
2) Calculate the weight as 
\begin{equation}
{w_{l,n}} =1/{  e}_{l,n}=    1 - \frac{{{{\left| {{\bf{h}}_n^H{{\bf{X}}_l}[n]} \right|}^2}}}{{\sum\nolimits_{i \ne n} {{{\left| {{\bf{h}}_n^H{{\bf{X}}_l}[i]} \right|}^2}}  + \sigma _n^2}}.
 \label{57_1}
\end{equation}  
  
 3) Using  \eqref{57_1} in \eqref{56_1},  calculate   the update of ${\bf X}_l$  by solving 
 \begin{equation} 
 \begin{aligned}
 &\mathop {\min}\limits_{    {\bf X}_l  }~  \sum\limits_{n = 1}^{N_U} \beta_n  {  w}_{l,n}  {  e}_{l,n} + \Re\left({\rm Tr}\left\lbrace{\bf D}_{1,l}^H \left(  {\bf X}_l-  {\bf F}_{\rm set} {\bf P} {\bf F}_{{\rm D},l} \right)  \right\rbrace    \right)\\
 &\qquad +\frac{\rho_{1}}{2}\left\|    {\bf X}_l-  {\bf F}_{\rm set} {\bf P} {\bf F}_{{\rm D},l} \right\|_F^2+ \Re\left({\rm Tr}\left\lbrace{\bf D}_{2,l}^H \left(  {\bf X}_l-  {\bf Z}_{l}  \right)     \right\rbrace \right)\\
 & \qquad +\frac{\rho_{2}}{2}\left\|    {\bf X}_l- {\bf Z}_l\right\|_F^2\\
 &  {\rm{s}}.{\rm{t}}.\quad     \sum\limits_{l = 1}^L  {\rm{Tr}}\left( {\bf X}_l {\bf{X}}_l^H \right) \le {\cal E}, 
 \end{aligned}\label{58}
 \end{equation}
whose closed form solution can be obtained similar to problem \eqref{42}. By introducing a Lagrange multiplier $ \mu $,  the first-order optimality condition of ${\bf X}_l$ is  
  \begin{equation} 
  \begin{aligned}
  & {\bf X}_l^{\rm opt}(\mu)= \left( {\bf \Xi}_l+\mu {\bf{I}}_{N_{\rm{Tx}}}\right) ^{ - 1}{\bf \Psi}_l,
  \end{aligned}\label{59_1}
  \end{equation}
  where ${\bf \Xi}_l$ and ${\bf \Psi}_l$ are defined as 
  \begin{equation} 
  \begin{aligned}
  {\bf \Xi}_l= \sum\limits_{n = 1}^{{N_U}} {{\beta _n}{w_{l,n}}{{\left| {{u_{l,n}}} \right|}^2}{{\bf{h}}_n}{\bf{h}}_n^H}  + \left( {\frac{{{\rho _1}}}{2} + \frac{{{\rho _2}}}{2}} \right){\bf{I}}_{N_{\rm{Tx}}}
  \end{aligned}\label{60_1}
  \end{equation}
  and 
  \begin{equation} 
  \begin{aligned}
  {\bf \Psi}_l= &  \sum\limits_{n = 1}^{{N_U}} {{\beta _n}{w_{l,n}}u_{l,n}^*{{\bf{h}}_n}{\bf{e}}_n^H}  - \frac{1}{2}\left( {{{\bf{D}}_{1,l}} + {{\bf{D}}_{2,l}}} \right) \\
  & + \frac{{{\rho _1}}}{2}{{\bf{F}}_{{\rm{set}}}}{\bf{P}}{{\bf{F}}_{{\rm{D}},l}} + \frac{{{\rho _2}}}{2}{{\bf{Z}}_l} 
  \end{aligned}\label{61}
  \end{equation}
with ${\bf e}_n$ is an $N_{\rm T_x}$ dimensional vector whose $n$-th entry is 1 and 0 otherwise. Then, the remaining procedure of the update of ${\bf X}_l$ is the same as Equations \eqref{46} and \eqref{48}.

  \vspace{-1em}
\section{Numerical Simulations}
This section provides various numerical simulations  
  to examine the performance of the proposed hybrid beamforming design for the DFRC system. We first
assess the  performance of the hybrid beamforming design with the THEREON algorithm for SU-MIMO scenario. Then,    the hybrid beamforming design for MU-MISO scenario is examined.

\begin{figure}[!tb]
	\centering
	\includegraphics[width=0.5\linewidth]{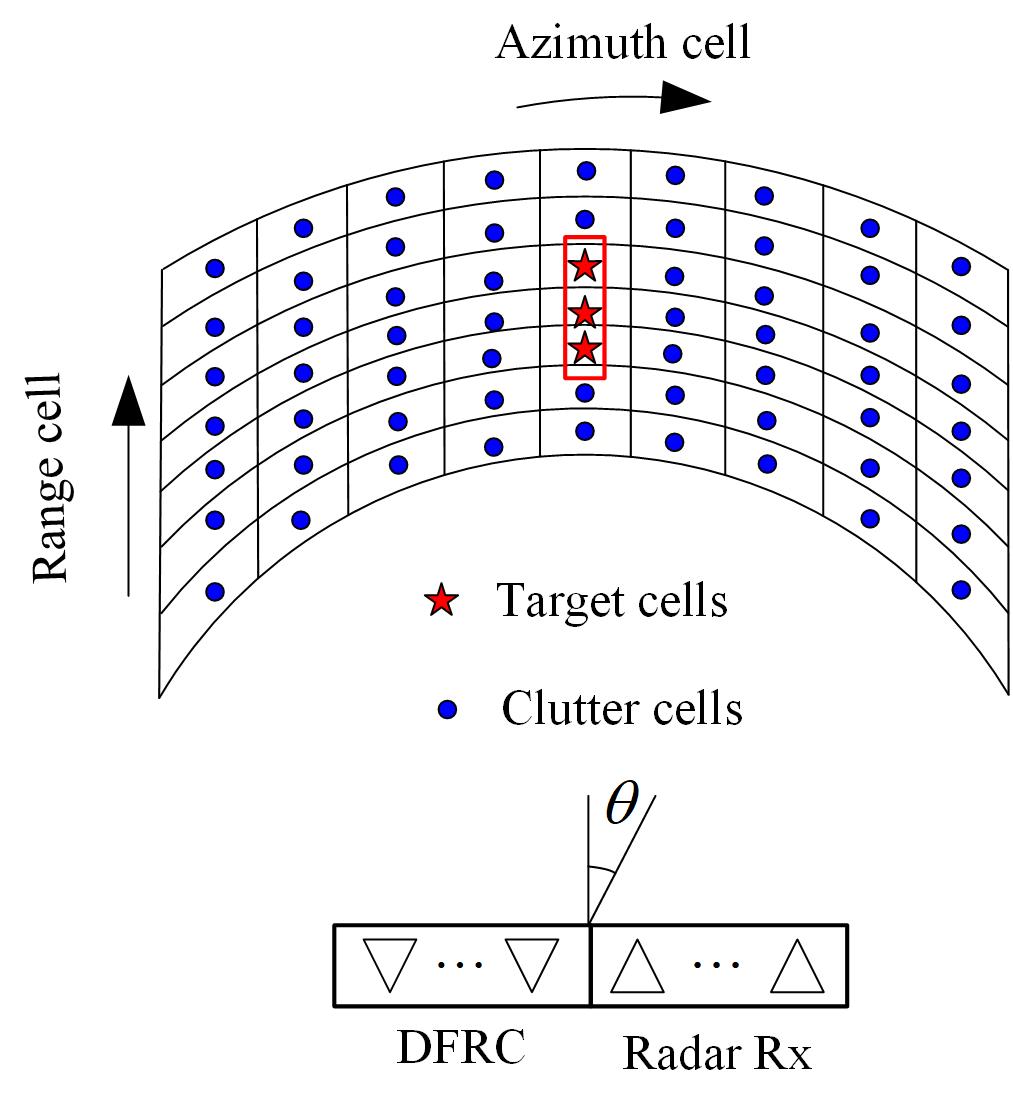}
	\caption{Range-azimuth cells of the illuminated area around the DFRC vehicle system. }
	\vspace{-0em}
	\label{fig:pic1_2}
\end{figure}

\begin{figure}[!tb]
	\centering
	\includegraphics[width=0.8\linewidth]{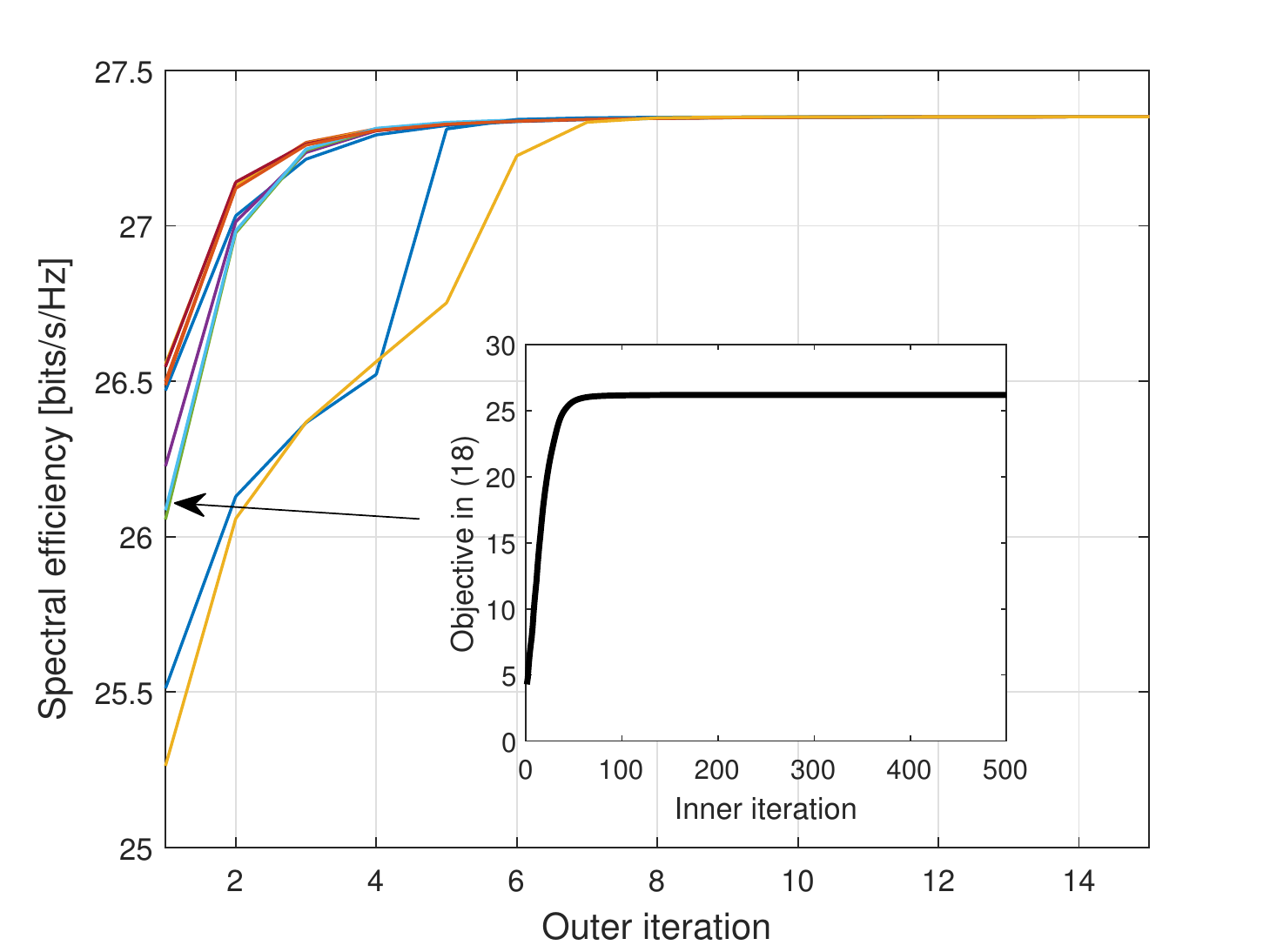}
	\caption{The convergence performance of the THEREON method for different
		initial points when considering  the intended SINR requirement   $\gamma=12 $, ${N_{\rm Tx}}=32$ and $N_{\rm RF}=4$.   }
	\label{fig:pic3}
\end{figure}

Unless  otherwise mentioned, in all simulations,
we assume a DFRC system with $N_{\rm Tx}= 32 $ transmit antennas.  The  radar receive array with $N_{\rm Rad}=4$ is considered. In simulations for the SU-MIMO case, the transmitter sends $N_s=4$ data symbols per subpulse to  a user   equipped with $2$ antennas. We assume
an communication environment with $N_{path}=16$ clusters, and the noises at users    are modelled as additive White Gaussian with   with the
covariances  of $\sigma_c^2=0.1$.

 For radar scenario, we assume that the Tx/Rx arrays boresight directions are used as the reference for the azimuth $\theta$, and that an extended target located at $ \theta_t=0^\circ $ (as illustrated in Fig. 2). The number of subpulses in each pulse is $L=16$, which is the case for $\mu=4$ in the recently published 5G NR standard \cite{dahlman20205g}. In this case, each subframe contains $16$ slots, and the length of each  slot is $62.5$ us. For modelling the impulse response of the extended target, we use the exponentially shaped covariance  to model the target second-order statistic matrix ${\bf \Sigma}_t ={\mathbb E}\{{\bf t}  {\bf t}^H \} $, that is, $ {\bf \Sigma}_t(m,n)=\sigma_{\alpha}^2 \eta_{\alpha}^{-|m-n|}, 1\le m,n\le L_{\rm tar} $ with $L_{\rm tar}=6 $, $\sigma_{\alpha}^2 = 10$ and $\eta_{\alpha}=15$. For the signal-dependent interference (i.e., clutters), we consider a homogeneous
 clutter environment composed of $ K=31$ azimuth cells, the azimuth angle of the $i$-th cell is $ \theta_i=2\pi(i-1)/K $. All  clutter second order statistic matrices $ {\bf \Sigma}_{c,i}={\mathbb E}\{{\bf j}_i  {\bf j}_i^H \} $ are identically modeled as ${\bf \Sigma}_t$ with $ {\bf \Sigma}_{c,i}(m,n)=\sigma_{\beta}^2 \eta_{\beta}^{-|m-n|}, \forall i, 1\le m,n\le L_{c,i} $ with $L_{c,i}=8 $, $\sigma_{\beta}^2 = 1$ and $\eta_{\beta}=1.2$.  As for   the radar receive  noise, we assume  corruption by a white  noise with the variance $\sigma_r^2=0.1$. Note that all numerical examples are analyzed using Matlab 2018b version and performed in a standard PC (with CPU Core i7 3.1 GHz and 16 GB RAM).

 \vspace{-1em}
\subsection{Hybrid Beamforming Design for SU-MIMO Scenario}

In the first example,  we  examine  the convergence performance of the proposed algorithm THEREON for solving problem  \eqref{16}. 
We consider the DFRC system with $N_{\rm RF}=4$ RF chains and the {\color{black}total energy} of the system is ${\cal E}=10$. The intended SINR requirement for the target is $\gamma=12 $ dB.   We set the penalty parameters as $\rho_1=\rho_2=20$.  Fig. \ref{fig:pic3}  analyzes the effect of 10 different initial points on
 the   convergence performance of the proposed THEREON framework  for solving
 problem \eqref{16}. For each initial point, we assume the entries of initial ${\bf F}_{\rm RF}$ are $e^{\jmath \Phi}$, where $\Phi$ obeys the uniform distribution over $(0, 2\pi]$ and entries of initial ${\bf F}_{\rm D}$ obey ${\cal CN}(0,1)$.  As shown in the figure,  the converged objective values are the same for different initial points can converge to the same value as the outer iteration (i.e. firstly update the radar filter $\bf V$ for given  $  ({{\bf F}_{\rm D}, {\bf F}_{\rm set} }, {\{{\bf U}_l\}}) $, and then update   $  ({{\bf F}_{\rm D}, {\bf F}_{\rm set} }, {\{{\bf U}_l\}}) $ with aid of the consensus-ADMM for given $\bf V$) goes on.  In addition, for  an instance of the THEREON framework, we also plot the convergence of  the objective values of problem \eqref{32} versus the inner iteration number by using the consensus-ADMM algorithm. The result  shows that  the objective value obtained by the consensus-ADMM  is able to  converge  to a sub-optimal value with the increasing iteration number. 

{\color{black}The performance of the consensus-ADMM with respect to maximum, minimum and average computation times until the termination condition are reached for different numbers of RF chains  is analyzed in Table I, where 100  Monto-Carlo trails are conducted.  The results show that the proposed consensus-ADMM has a good  computational efficiency.}

{
        \begin{table}[!t]
          \centering
          \fontsize{8}{11}\selectfont
          \caption{\color{black}Maximum, minimum and 
                average computation times (seconds) of the consensus-ADMM method}
          \label{tab:1}
             \begin{tabular}{|c|c|c|c|c|c|c|}
            \hline
            \multirow{1}{*}{$  N_{\rm RF} $ }&
               maximum time  &  minimum time &    average time    \cr
            \hline \hline
           2 &  1.14  & 1.35 & 1.28  \cr\hline
           4& 1.89 & 2.25 & 2.03  \cr\hline
           8& 2.83 & 3.46 & 3.11   \cr\hline
           12& 3.14 & 3.67 & 3.34   \cr\hline
            \end{tabular}
        \end{table}
        }

\begin{figure}[!tb]
	\centering
    \includegraphics[width=0.8\linewidth]{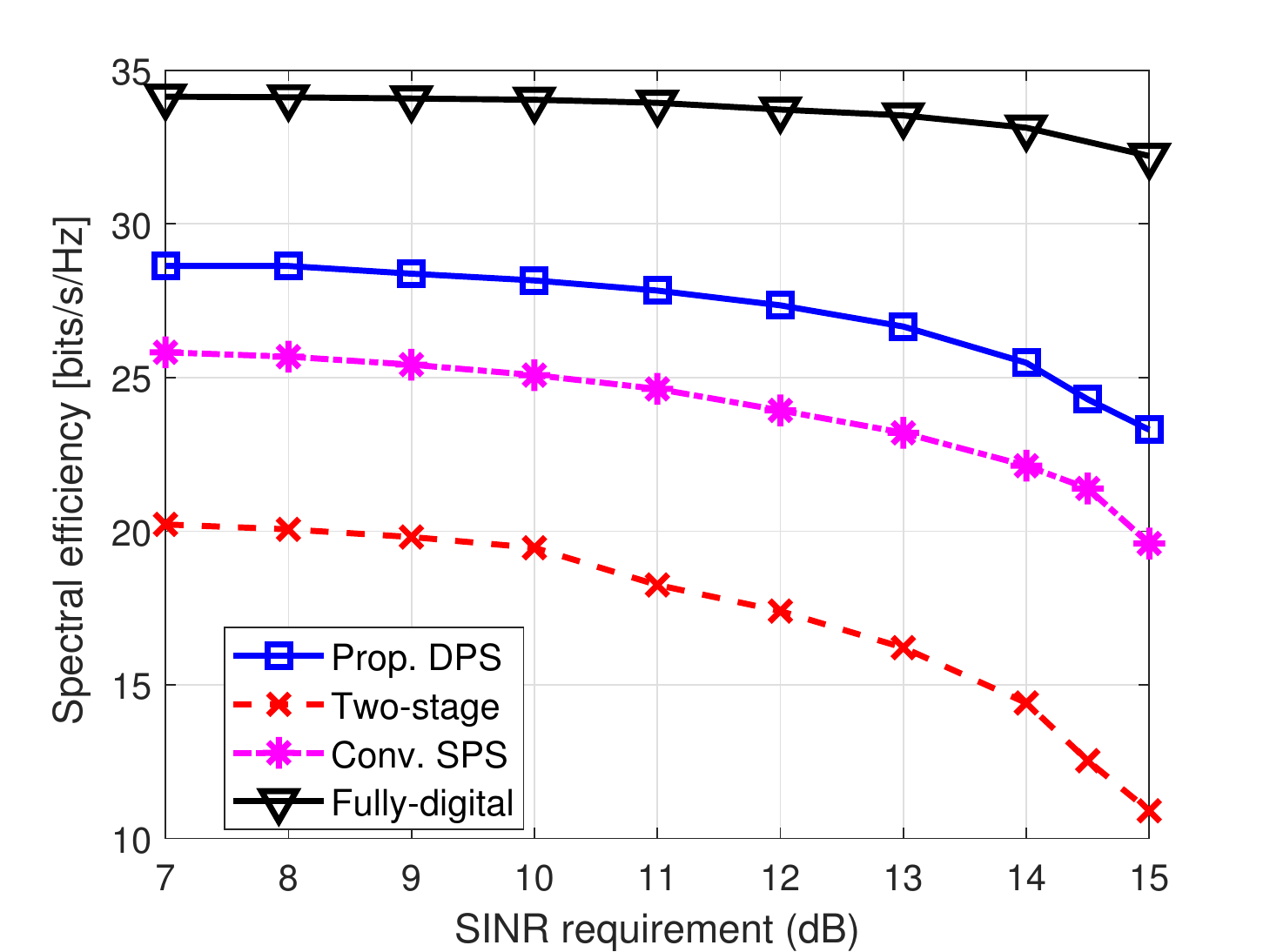}
	\caption{ The achieved SE values of different methods   versus  the intended radar SINR requirement $ \gamma $ when considering ${N_{\rm Tx}}=32$, $N_{\rm RF}=4$ and   ${\cal E}=10$. } 
	\label{fig:pic1_1}
    \vspace{-1em}
\end{figure}

Next, we  evaluate the performance of the proposed DPS hybrid beamforming design presented in Section IV for  the SU-MIMO system.  
Fig. \ref{fig:pic1_1} plots the SE value under the proposed DPS (denoted by ``Prop. DPS") architecture  versus  the intended radar SINR requirement $ \gamma $.  For comparison purpose, the fully-digital
beamformer (denoted by ``fully-digital")  which provides the upper-bound SE, the two-stage method (denoted by ``two-stage")  in  \cite{yu2019doubling,yu2016alternating}
and the conventional single phase shifter (SPS) architecture (denoted by ``Conv. SPS") are also considered.  The results show that the   obtained SE values  decrease along with the $ \gamma $, this is because when the intended  $\gamma$ is higher, the less degrees of freedom (DoFs)  can be used to maximize the communication SE. 
Thus there is a trade-off between the radar SINR behavior and communication performance. 
\textcolor{black}{
In addition, Fig. \ref{fig:pic1_1} also shows that both the proposed DPS and conventional SPS structures with the consensus-ADMM  achieve better SE values than the two-stage method in  \cite{yu2019doubling,yu2016alternating} which seeks to minimize the distance of the optimal fully-digital beamformers.  
}
Moreover, the proposed DPS achieves a better performance consistently over different radar SINR requirements with the SE gap of about 3 bps/Hz in comparison with the conventional SPS. 

\begin{figure}[!tb]
	\centering
    \includegraphics[width=0.8\linewidth]{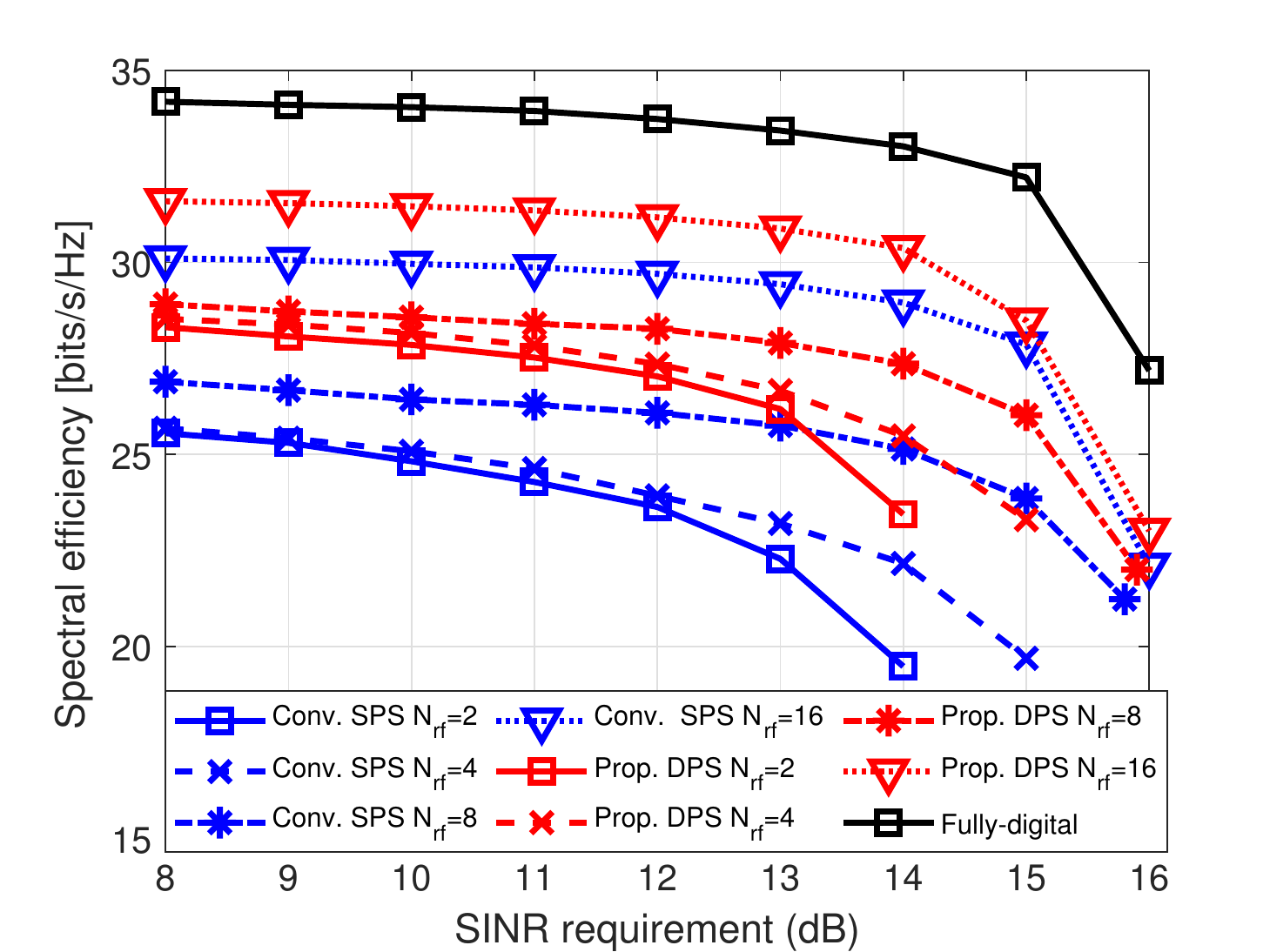}
	\caption{ The achieved SE   value   versus  the intended radar SINR requirement $ \gamma $  for for different numbers of RF chains when considering ${N_{\rm Tx}}=32$ and  ${\cal E}=10$.}
	\label{fig:pic4}
    \vspace{-1em}
\end{figure}

Fig. \ref{fig:pic4} displays the  SE  value versus  the intended SINR requirement $ \gamma $ for different numbers of RF chains $ N_{\rm RF}= 2, 4, 8, 16$ when considering ${N_{\rm Tx}}=32$ and the {\color{black}total energy} ${\cal E}=10$. {\color{black}For the case $N_{\rm RF}= 2$, the number of streams is considered to be 2}.    As expected,  the larger the number of RF chains, the higher the achieved communication   SE.    Besides, we also note that as the $ N_{\rm RF} $ increases, the gap between the proposed DPS and conventional SPS becomes smaller and smaller, and that the degradation trend of the  SE becomes larger and larger as the $\gamma$ increases. Furthermore,  Fig. \ref{fig:pic4} shows that when the $ N_{\rm RF} $ increases, achieving the radar SINR tends to be easier. {\color{black}This is because if the $ N_{\rm RF} $ is larger, the larger  the  degrees of freedom (DoFs) in the optimization design can be used to suppress  clutter, resulting the  better radar SINR.} This phenomenon agrees with our expectation.

\begin{figure}[!tb]
	\vspace{-1em}
	\centering
	\includegraphics[width=0.8\linewidth]{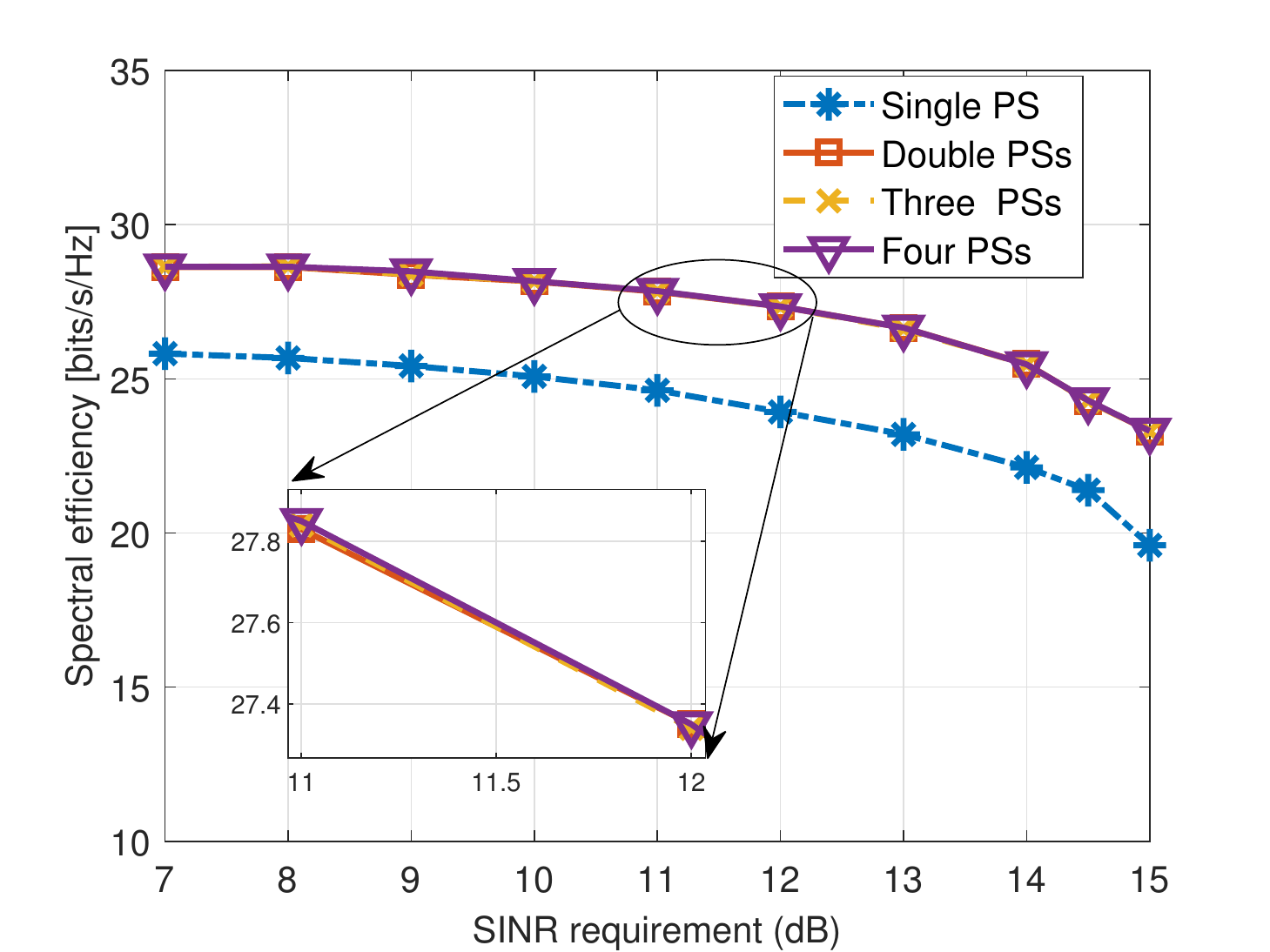}
    \vspace{-1em}
	\caption{The achieved SE values based on different PS structures for different radar SINR requirements with a fixed HBF power budget.}
	\label{fig:picNew}
    \vspace{-1em}
\end{figure}

Finally, we investigate the influence of adding more phase shifters into each DPS structure when the total HBF power budget is fixed. As shown in Fig. \ref{fig:picNew}, there are two points we can see clearly: (1) Comparing to the single PS case, the other cases have the extra performance gain  around $4$ dB constantly at different radar SINR level. (2) For all the PS structures with 2, 3, and 4 PS's, the corresponding curves are close to each other, where the differences are probably caused by the numerical accuracy. This indicates that, for a fixed power budget, the DPS structure is already capable to fully exploit the amplitude controlling in terms of improving the system performance.

\vspace{-1em}
\subsection{Hybrid Beamforming Design for MU-MISO Scenario}
\begin{figure}[!tb]
	\centering
    \includegraphics[width=0.8\linewidth]{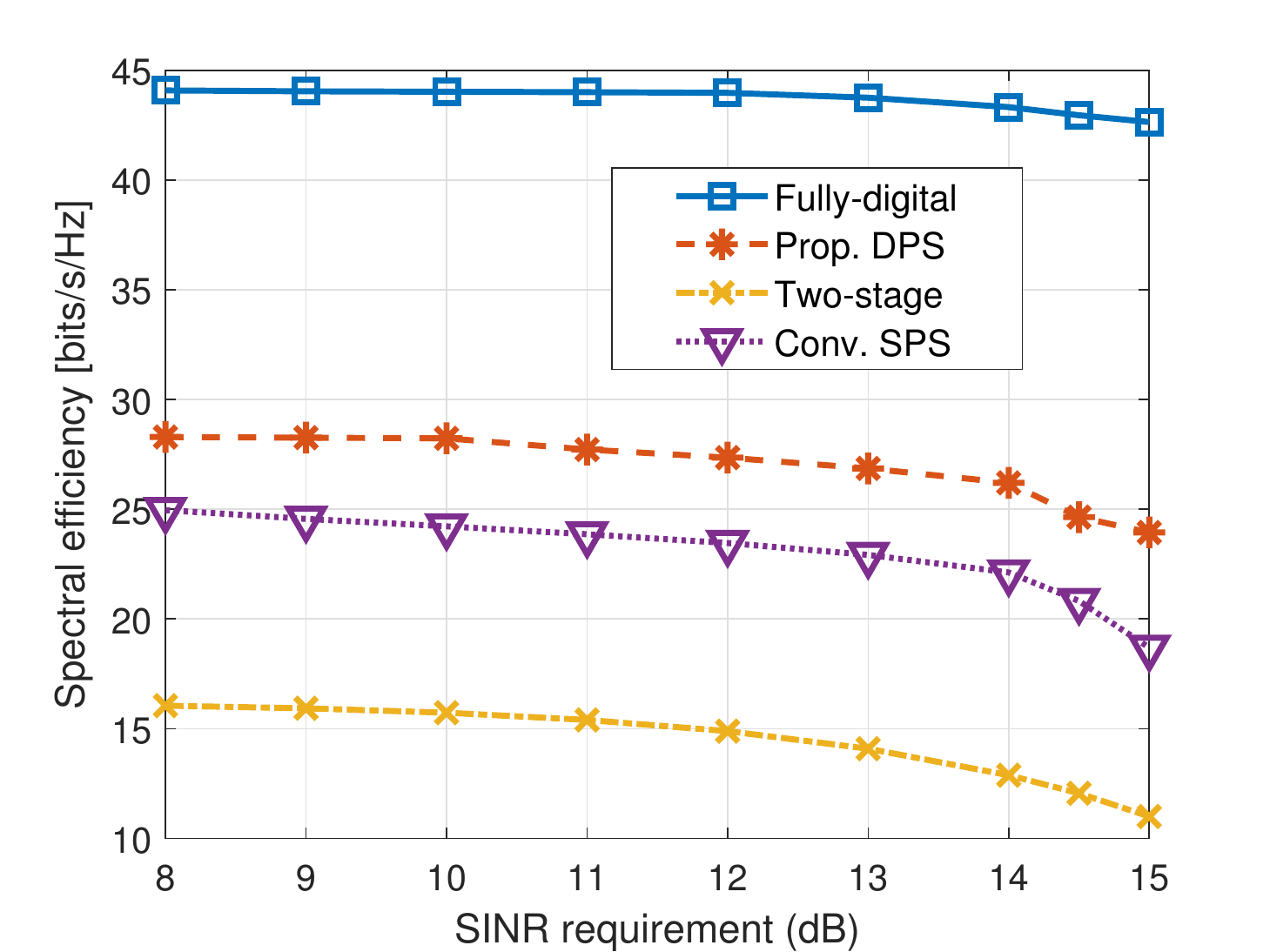}
    \caption{The sum spectral efficiency  obtained by different methods  versus the radar SINR requirement $\gamma$.  }
    \label{fig:pic6}
    \vspace{-1em}
\end{figure}

\begin{figure}[!tb]
	\centering
    \includegraphics[width=0.8\linewidth]{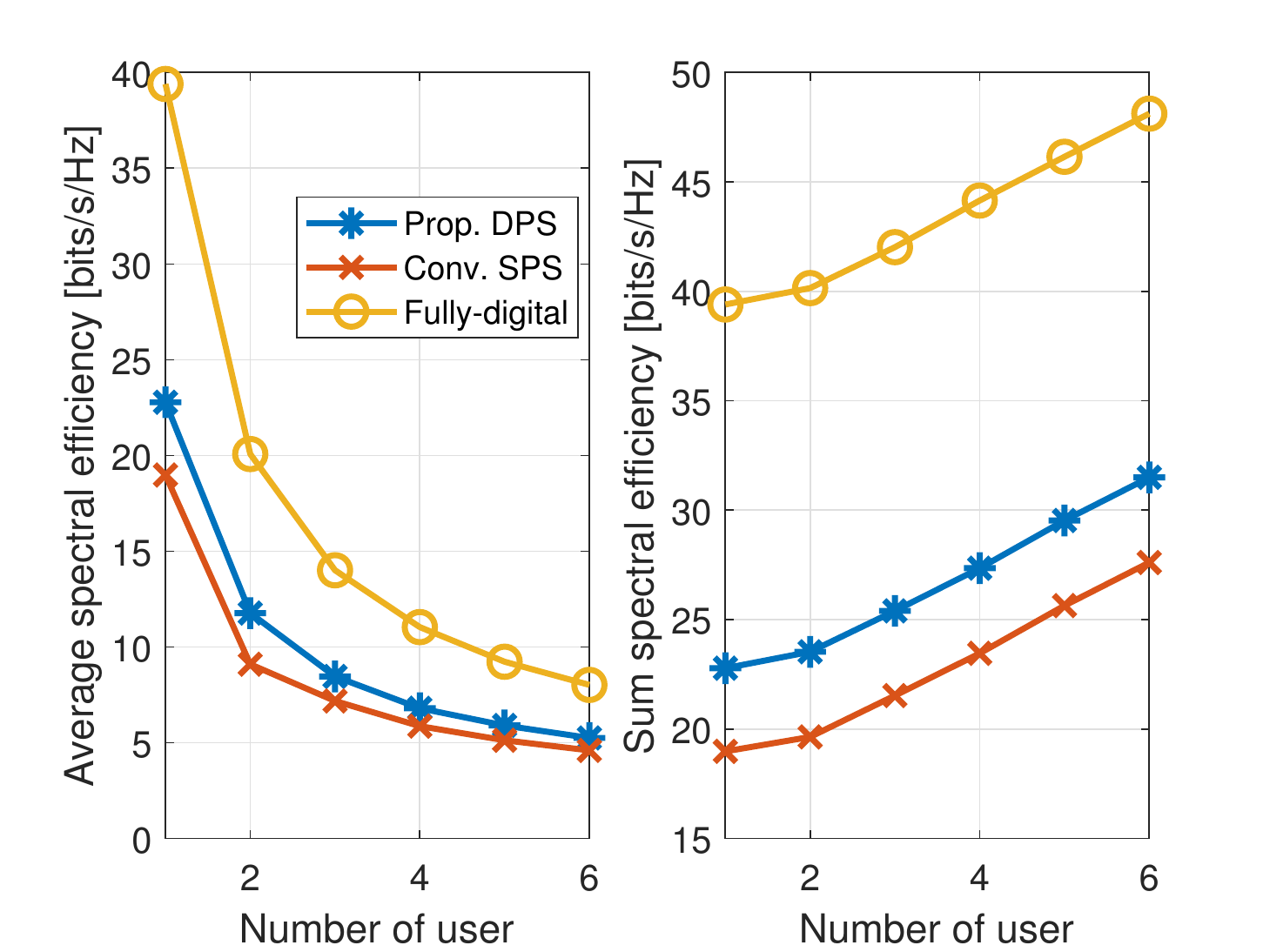}
    \caption{The   effect of number of users on the communication SE when considering $\gamma=12$ dB.}
    \label{fig:pic7}
    \vspace{-1em}
\end{figure}

In this subsection, we  assess   the proposed DPS hybrid beamforming design presented in Section IV for  the SU-MIMO system, in
which a DFRC vehicle  with $ 32 $ antennas  single-antenna  users and detects the target from stationary clutters environment simultaneously. We assume  that the priority weights of all users are set to be the same. 
Fig. \ref{fig:pic6} shows   SE obtained by different methods  versus the radar SINR requirement with  $N_{\rm RF}=4$ RF chains serves $N_U=4$. It can be seen that the proposed DPS hybrid beamforming method achieves much higher  SE than the conventional SPS and  the two-stage method in  \cite{8683591,yu2016alternating}. This implies  that the DPS structure  is  beneficial to improving the spectral efficiency. 
Finally, Fig. \ref{fig:pic7} analyzes the effect of number of users on the communication SE when considering $\gamma=12$ dB.  Specifically, we use the averaged SE (i.e., $S/U$) to describe system performance in left figure. As expected, the proposed DPS scheme outperforms the conventional SPS. In addition, it is interesting to note that as the number of users $ U$ increases, the averaged SE becomes
lower and lower.  This is because the larger $ U  $ means the MUI received by each use is stronger, which leads to the worse averaged SE. Besides, the sum SE is also shown in the right figure. From the figure, we find that    the sum SE increases with the number
 of users. 

\section{Conclusion}


This paper considers the problem of the DPS-based HBF design for the mmWave DFRC system in the presence of the extended target and clutters. By designing the HBF, we maximize the communication spectral efficiency while guarantee the predefined radar SINR level and the HBF power budget. To solve the formulated nonconvex problem, a low complexity method based on the consensus-ADMM approach is proposed to optimize the DPS-based HBF. In addition, we have extended the proposed method from the single-user scenario to the MU-MISO one. The simulation results demonstrate that the DPS structure improves the system performance with a moderate increase of phase shifters. Accordingly, the proposed HBF system achieves an superior trade-off between the radar and communication properties comparing to the conventional SPS architecture. 

Furthermore, we would like to emphasize that  the accurate information of the FIRs of target and clutters is assumed known in this work. This assumption can be practically relaxed by considering the uncertainty range of the information, and consequently, the robust design problem could be formulated in the form of the max-min. In addition, the partial connection architecture of the phase shifter network can be extended to the dynamic one, for which the DPS can still be applied so that the system performance is expected to be further enhanced. 
\appendices

 \section{Proof of  Proposition 1}
Based on the fact that ${\bf s}_l$ and $\bf t$   are statistically independent, we have
\begin{equation}
\begin{aligned}
& {\mathbb E}\left\{ {{{\left| {{\rm{Tr}}\left\{ {{{\bf{V}}^H}{{\bf{H}}_t}({\theta _t}){\bf{XT}}} {\cal F}_d\right\}} \right|}^2}} \right\} \\
& = {\mathbb E}\left\{ {{{\left| {{\rm{Tr}}\left\{ {{{\bf{V}}^H}{{\bf{H}}_t}({\theta _t}){\bf{XT}}} \right\}} \right|}^2}} \right\} \\
& ={\bf v}^H \Big({\bf I}_{L_{\rm obs}} \otimes  {{\bf{H}}_t}({\theta _t})\Big) {\mathbb E}\left\{ {\rm vec}\big( {\bf X T} \big) {\rm vec}^H\big( {\bf X T} \big)   \right\} \\
& \qquad \qquad \times \Big({\bf I}_{L_{\rm obs}} \otimes  {{\bf{H}}_t}({\theta _t})\Big)^H {\bf v},
\end{aligned}\label{66}
\end{equation}
where ${\bf v}={\rm vec}({\bf V})$. 
Given that ${\rm vec}({\bf X}{\bf T})=\widetilde{\bf X}{\bf t}$ with 
\[\small{\widetilde{\bf X}=\left[ {\begin{array}{*{20}{c}}
	{{\bf{x}}[1]}&{}&{\bf{0}}\\
	\vdots & \ddots &{}\\
	{{\bf{x}}[L]}& \ddots &{{\bf{x}}[1]}\\
	{}& \ddots &{}\\
	{\bf{0}}&{}&{{\bf{x}}[L]}
	\end{array}} \right]\in {\mathbb C}^{N_{\rm Tx}L_{\rm obs} \times L_{\rm tar}},}\]
where ${\bf x}[l]$ is the $l$-th column of $\bf X$, 
\eqref{66} can be written as 
\begin{equation}
\begin{aligned} & \mathbb{E}\left\{ \left|{\rm Tr}\left\{ {\bf V}^{H}{\bf H}_{t}(\theta_{t}){\bf XT}\right\} \right|^{2}\right\} \\
& = {\bf v}^{H}\Big({\bf I}_{L_{{\rm obs}}}\otimes{\bf H}_{t}(\theta_{t})\Big)\mathbb{E}_{{\bf s}}\left\{ \widetilde{{\bf X}}{\bf \Sigma}_{t}\widetilde{{\bf X}}^{H}\right\} \Big({\bf I}_{L_{{\rm obs}}}\otimes{\bf H}_{t}(\theta_{t})\Big)^{H}{\bf v}.
\end{aligned}
\end{equation}

Let $ \widetilde{\bf X}=[\tilde{\bf x}_1, \cdots, \tilde{\bf x}_{L_{\rm tar}}] $, we have 
  \begin{equation}
  \begin{aligned}
& {\mathbb E}_{\bf s}\left\{  \widetilde{\bf X} {\bf \Sigma}_t \widetilde{\bf X}^H    \right\}=  \sum\limits_{i = 1}^{{L_{{\rm{tar}}}}} {\sum\limits_{j = 1}^{{L_{{\rm{tar}}}}} {{{\bf{\Sigma }}_t}[i,j]} {\mathbb E}_{\bf s}\left\{ {{{\tilde {\bf{x}}}_i}\tilde {\bf{x}}_j^H} \right\}}.
 \end{aligned} 
 \end{equation}
Due to $ {\mathbb E}\{{\bf s}_l {\bf s}_l^H \} = {\bf I}_{N_s} $, one further gets 
\begin{equation}
 \begin{aligned}
&{\mathbb E}_{\bf s}\left\{ {{{\tilde {\bf{x}}}_i}\tilde {\bf{x}}_j^H} \right\} = {\bf{\Gamma }}_{ij}, 1\le i,j \le L_{\rm tar},
 \end{aligned} 
 \end{equation}
where  $ {\bf{\Gamma }}_{ij} \in {\mathbb C}^{N_{\rm Tx}  L_{\rm obs} \times N_{\rm Tx}L_{\rm obs} } $ and its  $ (n,m) $-th block is defined as 
\begin{equation}
{\bf \Gamma}_{ij}(n,m)=\begin{cases}
{\bf F}_{l}{\bf F}_{l}^{H} & n=i+l-1,m=j+l-1, \forall l,\\
{\bf 0}_{N_{{\rm Tx}}\times N_{{\rm Tx}}} & \text{otherwise}
\end{cases}
\end{equation}
with  ${\bf F}_l={\bf F}_{\rm RF} {\bf F}_{{\rm D},l} $.

By defining 
\begin{equation}
\begin{aligned}
&{{\bf{\Theta }}_t}({{\bf{F}}_{{\rm{RF}}}},{{\bf{F}}_{\rm{D}}}) \buildrel \Delta \over =  \\
& ({{\bf{I}}_{{L_{{\rm{obs}}}}}} \otimes {{\bf{H}}_t}({\theta _t}))\left( {\sum\limits_{i = 1}^{{L_{{\rm{tar}}}}} {\sum\limits_{j = 1}^{{L_{{\rm{tar}}}}} {{{\bf{\Sigma }}_t}[i,j]} {{\bf{\Gamma }}_{ij}}} } \right) {({{\bf{I}}_{{L_{{\rm{obs}}}}}} \otimes {{\bf{H}}_t}({\theta _t}))^H},
\end{aligned} 
\end{equation}
we obtain 
\begin{equation}
\begin{aligned}
& {\mathbb E}\left\{ {{{\left| {{\rm{Tr}}\left\{ {{{\bf{V}}^H}{{\bf{H}}_t}({\theta _t}){\bf{XT}}} \right\}} \right|}^2}} \right\}=
 {\bf v}^H{{\bf{\Theta }}_t}({{\bf{F}}_{{\rm{RF}}}},{{\bf{F}}_{\rm{D}}}) {\bf v}.
\end{aligned} 
\label{71_1}
\end{equation}

On the other hand, exploiting  the derivation similar to \eqref{66}, we have      
\begin{equation}
\begin{aligned}
& {\mathbb E} \left\{ {{{\left| {{\rm{Tr}}\left\{ {{{\bf{V}}^H}{{\bf{H}}_i}({\theta _i}){\bf{X}}{{\bf{J}}_i}} \right\}} \right|}^2}} \right\} \\
& =  {\bf v}^H \Big({\bf I}_{L_{\rm obs}} \otimes  {{\bf{H}}_i}({\theta _i})\Big)  {\mathbb E}\left\{ {\rm vec}\big( {{\bf X} {\bf J}_i} \big) {\rm vec}^H\big( {{\bf X} {\bf J}_i} \big)   \right\} \\
& \qquad \qquad \times \Big({\bf I}_{L_{\rm obs}} \otimes  {{\bf{H}}_i}({\theta _i})\Big)^H {\bf v}.
\end{aligned}\label{71}
\end{equation}

According to the equality that ${\rm vec}({\bf X}{\bf J}_i)=\widehat{\bf X}_i{\bf j}_i$ with 
\[{\widehat{\bf X}_i=\left[ {\begin{array}{*{20}{c}}
	{{\bf{x}}[1]}&{}&{\bf{0}}\\
	\vdots & \ddots &{}\\
	{{\bf{x}}[L]}& \ddots &{{\bf{x}}[1]}\\
	{}& \ddots &{}\\
	{\bf{0}}&{}&{{\bf{x}}[L]}
	\end{array}} \right]\in {\mathbb C}^{N_{\rm Tx}L_{\rm obs} \times L_{c,i}},}\]
and letting $ \widehat{\bf X}=[\hat{\bf x}_1,\cdots,\hat{\bf x}_{L_{c,i}}] $, we have 
\begin{equation}
\begin{aligned}
& {\mathbb E}_{\bf s}\left\{  \widehat{\bf X} {\bf \Sigma}_{c,i} \widehat{\bf X}^H    \right\}=  \sum\limits_{k = 1}^{{L_{c,i}}} {\sum\limits_{p= 1}^{{L_{c,i}}} {{{\bf{\Sigma }}_{c,i}}[p,k]} {\mathbb E}_{\bf s}\left\{ {{{\tilde {\bf{x}}}_p}\tilde {\bf{x}}_k^H} \right\}} ,
\end{aligned} 
\end{equation}
where 
\begin{equation}
\begin{aligned}
&{\mathbb E}_{\bf s}\left\{ {{{\hat {\bf{x}}}_p}\hat {\bf{x}}_k^H} \right\} =  {{{\bf{\Gamma }}_{pk}^{c,i}   }} \in   {\mathbb C}^{N_{\rm Tx}  L_{\rm obs} \times N_{\rm Tx}L_{\rm obs} }
\end{aligned} 
\end{equation}
with the $ (n,m) $-th block of the matrix $ {\bf{\Gamma }}_{pk}^{c,i}  $ defined as 
\begin{equation}
 \begin{aligned}
  &  {\bf{\Gamma }}_{pk}^{c,i} (n, m )= \left\{ \begin{array}{l}
  {{\bf{F}}_{l}}{\bf{F}}_{l}^H,\;\;  ~~n= p+l-1,m=k+l-1, \forall l\\
  {{\bf{0}}_{{N_{{\rm{Tx}}}} \times {N_{{\rm{Tx}}}}}},\;\;\;\;  {\rm{otherwise}}
  \end{array} \right.
 \end{aligned}
\end{equation}

Define 
\begin{equation}
\begin{aligned}
& {\bf \Theta}_{c,i}({\bf F}_{{\rm RF}},{\bf F}_{{\rm D}}) \triangleq \\
&   {\bf I}_{L_{{\rm obs}}}\otimes{\bf H}_{i}(\theta_{i}))\left(\sum\limits _{k=1}^{L_{c,i}}\sum\limits _{l=1}^{L_{c,i}}{\bf \boldsymbol{\Sigma}}_{c,i}[l,k]\boldsymbol{{\bf \Gamma}}_{lk}^{c,i}\right)({\bf I}_{L_{{\rm obs}}}\otimes{\bf H}_{i}(\theta_{i}))^{H}
\end{aligned}
\end{equation}
we obtain 
\begin{equation}
\begin{aligned}
& {\mathbb E} \left\{ {{{\left| {{\rm{Tr}}\left\{ {{{\bf{V}}^H}{{\bf{H}}_i}({\theta _i}){\bf{X}}{{\bf{J}}_i}} \right\}} \right|}^2}} \right\} =
{\bf v}^H{{\bf{\Theta }}_{c,i}}({{\bf{F}}_{{\rm{RF}}}},{{\bf{F}}_{\rm{D}}}) {\bf v}
\label{76}
\end{aligned} 
\end{equation}

Additionally,  let ${\bf z}_r={\rm vec} ({\bf Z}_r)$, one gets 
\begin{equation}
{\mathbb E}\left\{ {{{\left| {{\rm{Tr}}\left\{ {{{\bf{V}}^H}{{\bf{Z}}_r}} \right\}} \right|}^2}} \right\} ={\bf v}^H   {\mathbb E}\left\{{\bf z}_r {\bf z}_r^H     \right\}   {\bf v} =\sigma_r^2 {\bf v}^H   {\bf v} 
\label{77}
\end{equation}
Based on \eqref{71_1}, \eqref{76} and \eqref{77},  we can attain that 
\begin{equation}
 {\rm SINR}({\bf F}_{\rm RF}, {\bf F}_{D}, {\bf V}) =\dfrac{{\bf v}^H{{\bf{\Theta }}_t}({{\bf{F}}_{{\rm{RF}}}},{{\bf{F}}_{\rm{D}}}) {\bf v}}{{\bf v}^H{{\bf{\Theta }}_{c}}({{\bf{F}}_{{\rm{RF}}}},{{\bf{F}}_{\rm{D}}}) {\bf v}+ \sigma_r^2 {\bf v}^H   {\bf v}   }
\end{equation}
where $  {{\bf{\Theta }}_{c}}({{\bf{F}}_{{\rm{RF}}}},{{\bf{F}}_{\rm{D}}}) =  \sum\limits_{i= 1}^K   {{\bf{\Theta }}_{c, i}}({{\bf{F}}_{{\rm{RF}}}},{{\bf{F}}_{\rm{D}}})   $.

Next, let us derive \eqref{14b}, since 
\begin{equation}
\begin{aligned} 
& {\mathbb{E}}\left\{ {{\left|{{\rm {Tr}}\left\{ {{{\bf {V}}^{H}}{{\bf {H}}_{t}}({\theta_{t}}){\bf {XT}}}\right\} }\right|}^{2}}\right\} = {\mathbb{E}}\left\{ {{\left|{{\rm {Tr}}\left\{ {{{\bf {X}}^{H}}{{\bf {H}}_{t}^{H}}({\theta_{t}}){\bf V}{\bf {T}}^{H}}\right\} }\right|}^{2}}\right\} \\
& =  {\rm {Tr}}\Big\{ \Big({\bf I}_{L}\otimes{{\bf {H}}_{t}^{H}}({\theta_{t}})\Big){\mathbb{E}}\left\{ {\rm vec}\big({\bf V}{\bf T}^{H}\big){\rm vec}^{H}\big({\bf V}{\bf T}^{H}\big)\right\} \\
& \qquad \qquad \times \Big({\bf I}_{L}\otimes{{\bf {H}}_{t}}({\theta_{t}})\Big) {\mathbb{E}}_{{\bf s}}\left\lbrace {\bf x}{\bf x}^{H}\right\rbrace \Big\}.
\end{aligned}
\label{79}
\end{equation}
Utilizing the equality that ${\rm vec}({\bf V}{\bf T}^H)=\widetilde{\bf V}{\bf t}^*$ with 
\[\widetilde{\bf V}=\left[ {\begin{array}{*{20}{c}}
	{{\bf{v}}[1]}&{{\bf{v}}[2]}& \cdots &{{\bf{v}}[{L_{{\rm{tar}}}}]}\\
	{{\bf{v}}[2]}&{{\bf{v}}[3]}& \cdots &{{\bf{v}}[{L_{{\rm{tar}}}} + 1]}\\
	\vdots & \vdots & \vdots & \vdots \\
	{{\bf{v}}[L]}&{{\bf{v}}[L + 1]}& \cdots &{{\bf{v}}[{L_{{\rm{obs}}}}]}
	\end{array}} \right]\in {\mathbb C}^{N_{\rm Rad}L \times L_{\rm tar}}, \]
we have 
\begin{equation}
 {\mathbb E}\left\{ {\rm vec}\big( {\bf V}{\bf T}^H \big) {\rm vec}^H\big( {\bf V}{\bf T}^H \big)   \right\} =\widetilde{\bf V} {\bf \Sigma}_t \widetilde{\bf V}^H.
 \label{80}
\end{equation}
Besides,  since $ {\mathbb E}\{{\bf s}_l {\bf s}_l^H \} = {\bf I}_{N_s} $, one obtains 
\begin{equation}
 {\mathbb E}_{\bf s} \left\lbrace  {\bf x}{\bf x}^H  \right\rbrace = {\rm Bdiag} \left( {\bf F}_1{\bf F}^H_1,   \cdots,  {\bf F}_L{\bf F}^H_L \right)  \buildrel \Delta \over = \widetilde{\bf F}.
 \label{81}
\end{equation}
Plugging \eqref{80} and \eqref{81} into \eqref{79} yields 
\begin{equation}
\begin{aligned} 
& {\mathbb{E}}\left\{ {{\left|{{\rm {Tr}}\left\{ {{{\bf {V}}^{H}}{{\bf {H}}_{t}}({\theta_{t}}){\bf {XT}}}\right\} }\right|}^{2}}\right\} 
= {\rm {Tr}}\Big\{\widetilde{{\bf F}}{\bf \Phi}_{t}({\bf V})\Big\} \\
& =\sum\limits _{l=1}^{L}{{\rm {Tr}}\left\{ {{{\bf {F}}_{l}}{\bf {F}}_{l}^{H}{{\bf {\Phi}}_{t}}[l,l]}\right\} },
\end{aligned}
\label{82}
\end{equation}
where $  {\bf \Phi}_t ({\bf V})  $ is defined as 
\begin{equation}
\begin{aligned}
{\bf \Phi}_t ({\bf V})&=\Big({\bf I}_{L} \otimes  {{\bf{H}}_t^H}({\theta _t})\Big)     \widetilde{\bf V} {\bf \Sigma}_t \widetilde{\bf V}^H 
  \Big({\bf I}_{L} \otimes  {{\bf{H}}_t}({\theta _t})\Big) \\
  & \buildrel \Delta \over =\left[ {\begin{array}{*{20}{c}}
  	{{{\bf{\Phi }}_t}[1,1]}& \cdots &{{{\bf{\Phi }}_t}[1,L]}\\
  	\vdots & \ddots & \vdots \\
  	{{{\bf{\Phi }}_t}[L,1]}& \cdots &{{{\bf{\Phi }}_t}[L,L]}
  	\end{array}} \right].
\end{aligned} \label{83}
\end{equation}

Finally, for $  {\mathbb E}\left\{ {{{\left| {{\rm{Tr}}\left\{ {{{\bf{V}}^H}{{\bf{H}}_i}({\theta _i}){\bf{X } \bf J}_i } \right\}} \right|}^2}}  \right\}   $, we have 
\begin{equation}
\begin{aligned} & {\mathbb{E}}\left\{ {{\left|{{\rm {Tr}}\left\{ {{{\bf {V}}^{H}}{{\bf {H}}_{i}}({\theta_{i}}){\bf {X}{\bf J}}_{i}}\right\} }\right|}^{2}}\right\} \\
& = {\rm {Tr}}\Big\{\widetilde{{\bf F}}{\bf \Phi}_{c,i}({\bf V})\Big\}=\sum\limits _{l=1}^{L}{{\rm {Tr}}\left\{ {{{\bf {F}}_{l}}{\bf {F}}_{l}^{H}{{\bf {\Phi}}_{c,i}}[l,l]}\right\} }
\end{aligned}
\label{84}
\end{equation}
where $  {\bf \Phi}_{c,i} ({\bf V})  $ is defined as 
\begin{equation}
\begin{aligned}
{\bf \Phi}_{c,i} ({\bf V})&=\Big({\bf I}_{L} \otimes  {{\bf{H}}_i^H}({\theta _i})\Big)     \widehat{\bf V}_i {\bf \Sigma}_{c,i} \widehat{\bf V}_i^H 
\Big({\bf I}_{L} \otimes  {{\bf{H}}_i}({\theta _i}) \Big) \\
& \buildrel \Delta \over =\left[ {\begin{array}{*{20}{c}}
	{{{\bf{\Phi }}_{c,i}}[1,1]}& \cdots &{{{\bf{\Phi }}_{c,i}}[1,L]}\\
	\vdots & \ddots & \vdots \\
	{{{\bf{\Phi }}_{c,i}}[L,1]}& \cdots &{{{\bf{\Phi }}_{c,i}}[L,L]}
	\end{array}} \right]
\end{aligned} \label{85}
\end{equation}
with  $   \widehat{\bf V}_i $ being given by 
\begin{equation}
    \widehat{\bf V}_i=\left[ {\begin{array}{*{20}{c}}
	{{\bf{v}}[1]}&{{\bf{v}}[2]}& \cdots &{{\bf{v}}[{L_{c,i}}]}\\
	{{\bf{v}}[2]}&{{\bf{v}}[3]}& \cdots &{{\bf{v}}[{L_{{{c,i}}}} + 1]}\\
	\vdots & \vdots & \vdots & \vdots \\
	{{\bf{v}}[L]}&{{\bf{v}}[L + 1]}& \cdots &{{\bf{v}}[{L_{{\rm{obs}}}}]}
	\end{array}} \right]\in {\mathbb C}^{N_{\rm Rad}L \times L_{{c,i}}}, 
\end{equation}

Based on \eqref{82} and \eqref{84}, we can obtain 
\begin{equation}
{\rm SINR}({\bf F}_{\rm RF}, {\bf F}_{D}, {\bf V}) =\dfrac{\sum\limits_{l = 1}^L {{\rm{Tr}}\left\{ {{{\bf{F}}_l}{\bf{F}}_l^H{{\bf{\Phi }}_t}[l,l]} \right\}} }{ \sum\limits_{l = 1}^L {{\rm{Tr}}\left\{ {{{\bf{F}}_l}{\bf{F}}_l^H{{\bf{\Phi }}_{c}}[l,l]} \right\}} + \sigma_r^2 {\bf v}^H   {\bf v}   }
\end{equation}
with $ {\bf \Phi}_{c} ({\bf V})  = \sum\limits_{i= 1}^K  {\bf \Phi}_{c,i} ({\bf V}) $. Thereby, this proof is completed. 

\vspace{-1em}
 {\color{black}\section{Proof of Theorem 1}
 The proof  mainly includes two steps.
 
i) For fixed ${\bf X}_l$,  the function $ f({\bf W}_l, {\bf X}_l, {\bf U}_l )$ is convex with respect to ${\bf U}_l$ and ${\bf W}_l$. Then the closed-form solutions
of ${\bf U}_l$ and ${\bf W}_l$ can be obtained by taking their first-order optimality conditions, which are given by 
$ {\bf U}_l^\star=\left(     {\bf H}   {\bf X}_l {\bf X}_l^H    {\bf H}^H +   \sigma_c^2 {\bf I}_{\rm Rx}   \right)^{-1}  {\bf H}   {\bf X}_l,$
and 
${\bf W}_l^\star= {\bf E}_l^{-1}({\bf X}_l, {\bf U}_l^\star ) = \left({\bf I}_{N_s}- {\bf X}_l^H {\bf H}^H   {\bf U}_l^\star  \right)^{-1}.$

ii) Substituting   ${\bf U}_l^\star$ and  ${\bf W}_l^\star$  into   $\min f({\bf W}_l, {\bf X}_l, {\bf U}_l )$ , we have 
  \begin{equation}\small
  \begin{aligned}
    & \min f({\bf W}_l^\star, {\bf X}_l, {\bf U}_l^\star )=   \max \sum\limits_{l = 1}^L  \log\Big|  {\bf W}_l^\star \Big| \\
    & =\max \sum\limits_{l = 1}^L  \log\Big| \left({\bf I}_{N_s}- {\bf X}_l^H {\bf H}^H   {\bf U}_l^\star  \right)^{-1} \Big|\\
    & =\max \sum\limits_{l = 1}^L  \log\Big| \Big({\bf I}_{N_s} - {\bf X}_l^H {\bf H}^H  \left(     {\bf H}   {\bf X}_l {\bf X}_l^H    {\bf H}^H +   \sigma_c^2 {\bf I}_{\rm Rx}   \right)^{-1}{\bf H}   {\bf X}_l  \Big)^{-1} \Big|. 
  \end{aligned}
  \end{equation}
  
  Let  $ {\bf Q}_1 \left( \begin{array}{ll}
 	{\bf{\Sigma }}^T , & 
 	{\bf{0}}^T
 	\end{array} \right)^T {\bf Q}_2^H  $  be the eigen-decomposition of ${\bf H}{\bf X}_l$, one gets 
\begin{equation}
    \begin{aligned}
     & \Big| \Big({\bf I}_{N_s}- {\bf X}_l^H {\bf H}^H  \left(     {\bf H}   {\bf X}_l {\bf X}_l^H    {\bf H}^H +   \sigma_c^2 {\bf I}_{\rm Rx}   \right)^{-1} 
   {\bf H}   {\bf X}_l  \Big)^{-1} \Big| \\
   & = \Big| \left({\bf I}_{N_s}- {\bf Q}_2   {{{\bf{\Sigma }}^T}{{\left( {{\bf{\Sigma }}{{\bf{\Sigma }}^T} + \sigma _c^2{\bf{I}}} \right)}^{ - 1}}{\bf{\Sigma }}}   {\bf Q}_2^H\right)^{ - 1} \Big| = \Big|    {\bf I}_{\rm RX}  +  {\bf \Sigma} {\bf \Sigma} ^T /\sigma_c^2     \Big|
    \end{aligned}\label{78}
\end{equation}
  
  On the other hand, substituting   ${\bf U}_l^\star$   into  $ R_l({\bf X}_l, {\bf U}_l )$ yields
\begin{equation}
\begin{aligned} 
&  R_{l}  ({\bf X}_{l},{\bf U}_{l}) 
 = \log\Big|{\bf I}_{N_{{\rm Rx}}}+{\bf U}_{l}{\bf C}_{l}^{-1}{\bf U}_{l}^{H}{\bf H}{\bf X}_{l}{\bf X}_{l}^{H}{\bf H}^{H}\Big| \\
 & =\log\Big| {\bf C}_{l}^{-1} \Big({\bf C} + {\bf U}_{l}^{H}{\bf H}{\bf X}_{l}{\bf X}_{l}^{H}{\bf H}^{H} {\bf U}_{l} \Big)\Big| \\
 & = \log\Big|{\bf C}_{l}^{-1}  {\bf U}_{l}^{H}  \big( \sigma_c^2 {\bf I}_{\rm Rx} + {\bf H}{\bf X}_{l}{\bf X}_{l}^{H}{\bf H}^{H}   \big)   {\bf U}_{l}  \Big| \\
 & = \log\Big|{\bf C}_{l}^{-1}  {\bf X}_{l}^{H}{\bf H}^{H}  \big( \sigma_c^2 {\bf I}_{\rm Rx} + {\bf H}{\bf X}_{l}{\bf X}_{l}^{H}{\bf H}^{H}   \big)^{-1}   {\bf H}{\bf X}_{l} \Big|\\
 & = \log\Big|\big(\sigma_c^2 {\bf Q}_2 {\bf{\Sigma }}^T ( \sigma_c^2{\bf I}_{N_{{\rm Rx}}}+  {\bf{\Sigma }}{\bf{\Sigma }}^T )^{-2} {\bf{\Sigma }} {\bf Q}_2^H \big)^{-1} \\
 & \qquad \qquad \times {\bf Q}_2  {\bf \Sigma}^T  \left( \sigma_c^2 {\bf I}_{\rm Rx} +  {\bf \Sigma} {\bf \Sigma} ^T \right)^{-1}      {\bf \Sigma}  {\bf Q}_2^H \Big| \\
 & =\left|  {\bf I}_{\rm Rx} +  {\bf \Sigma} {\bf \Sigma} ^T /\sigma_c^2   \right|.
\end{aligned}\label{79_3}
\end{equation}
According to \eqref{78} and \eqref{79_3}, this proof is completed. 
}

\section{Proof of  Theorem 2}
Specifically, introducing a Lagrange multiplier $ \mu $ on the energy constraint in problem \eqref{42}, we obtain the following Lagrangian function:
\begin{equation}
\begin{aligned}
 {\cal L}_x= &{\rm Tr}\left\lbrace {\bf E}_l ({\bf X}_l, {\bf U}_l )  {\bf W}_l \right\rbrace  + \Re\left({\rm Tr}\left\lbrace{\bf D}_{1,l}^H \left(  {\bf X}_l-  {\bf F}_{\rm set} {\bf P} {\bf F}_{{\rm D},l} \right)  \right\rbrace    \right) \\
  & +\frac{\rho_{1}}{2} \left\|    {\bf X}_l-  {\bf F}_{\rm set} {\bf P} {\bf F}_{{\rm D},l} \right\|_F^2 
  + \Re\left({\rm Tr}\left\lbrace{\bf D}_{2,l}^H \left(  {\bf X}_l-  {\bf Z}_{l}  \right)     \right\rbrace \right) \\
  & +\frac{\rho_{2}}{2}\left\|    {\bf X}_l- {\bf Z}_l\right\|_F^2 
  +\mu\Big(   \sum\limits_{l = 1}^L  {\rm{Tr}}\left( {\bf X}_l {\bf{X}}_l^H \right) -{\cal E} \Big).
\end{aligned}\label{79_1}
\end{equation}
whose first-order optimality condition is given by 
\begin{equation}
\begin{aligned}
& {\bf X}_l^{\rm opt}(\mu)=\left( {\bf \Xi}_l+\mu {\bf{I}}_{N_{\rm{Tx}}}\right) ^{ - 1}{\bf \Psi}_l,
\end{aligned}\label{43}
\end{equation}
where ${\bf \Xi}_l$ and ${\bf \Psi}_l$ are defined as 
${\bf \Xi}_l= {{\bf{H}}^H}{{\bf{U}}_l}{{\bf{W}}_l}{\bf{U}}_l^H{\bf{H}} + \left( {\frac{{{\rho _1}}}{2} + \frac{{{\rho _2}}}{2}} \right){\bf{I}}_{N_{\rm{Tx}}},$
and 
${\bf \Psi}_l= {{{\bf{H}}^H}{{\bf{U}}_l}{{\bf{W}}_l}}$
$ - \frac{1}{2}\left( {{{\bf{D}}_{1,l}} + {{\bf{D}}_{2,l}}} \right) {+ \frac{{{\rho _1}}}{2}{{\bf{F}}_{{\rm{set}}}}{\bf{P}}{{\bf{F}}_{{\rm{D}},l}} + \frac{{{\rho _2}}}{2}{{\bf{Z}}_l}},$

{\color{black}
Based on the complementary
slackness of the KKT, i.e., $\mu\Big(   \sum\limits_{l = 1}^L  {\rm{Tr}}\left( {\bf X}_l {\bf{X}}_l^H \right) -{\cal E} \Big)=0$, we have the following two cases:

i) If $\mu=0$, we attain the optimal ${\bf X}_l$ as 
$ {\bf X}_l^{\rm opt}(0)={\bf \Xi}_l^{ - 1} {\bf \Psi}_l$,
which must satisfy the condition $ \sum\limits_{l = 1}^L  {\rm{Tr}}\left( {\bf X}_l {\bf{X}}_l^H \right) \le {\cal E}$. }

ii) Otherwise, we must have $ \sum\limits_{l = 1}^L  {\rm{Tr}}\left( {\bf X}_l {\bf{X}}_l^H \right) = {\cal E}$. For this case, we define $ {\bf \Xi}_l= {\bf Q}_l {\bf \Lambda}_l {\bf Q}_l^H  $ be the EVD of  $ {\bf \Xi}_l$ and $\tilde{\bf \Psi}_l= {\bf Q}_l^H  {\bf \Psi}_l$,
we have 
\begin{equation}
\begin{aligned}
{\bf X}_l^{\rm opt}(\mu)={\bf Q}_l  \left( {\bf \Lambda}_l+\mu {\bf{I}}_{N_{\rm{Tx}}}\right) ^{ - 1}\tilde{\bf \Psi}_l.
\end{aligned}\label{46}
\end{equation}

Substituting \eqref{46} into the total power constraint  in \eqref{42}, we have
\begin{equation}
 \begin{aligned}
& \sum\limits_{l = 1}^L  {\rm{Tr}}\left( \left( {\bf \Lambda}_l+\mu {\bf{I}}_{N_{\rm{Tx}}}\right) ^{ - 1}\tilde{\bf \Psi}_l \tilde{\bf \Psi}_l^H   \left( {\bf \Lambda}_l+\mu {\bf{I}}_{N_{\rm{Tx}}}\right) ^{ - 1}  \right) \\
& \qquad =   \sum\limits_{l = 1}^L   \sum\limits_{n = 1}^{N_{\rm Tx}}   \dfrac{\left( \tilde{\bf \Psi}_l \tilde{\bf \Psi}_l^H\right)  [n,n]}{ \left({\bf \Lambda}_l[n,n]+\mu    \right)^2 }  ={\cal E},
  \end{aligned}
  \label{48}
  \end{equation}
where $  \left( \tilde{\bf \Psi}_l \tilde{\bf \Psi}_l^H\right)  [n,n] $ and $ {\bf \Lambda}_l[n,n] $ denote the $ (n,n) $-th element of $ \tilde{\bf \Psi}_l \tilde{\bf \Psi}_l^H  $ and ${\bf \Lambda}_l $, respectively.  
Since the left-hand side (LHS) of \eqref{48} is a decreasing function of $\mu$, and the unique solution $\mu^\star$ can be found by the bisection method \cite{boyd2004convex}.

 \section{Proof of  Theorem 3}
 More concretely, we introduce a dual variable  $\nu \ge 0$ for the constraint in problem \eqref{50}.   Based on the complementary slackness of the KKT, i.e. $ \nu\left( \sum\limits_{l = 1}^L  {\rm{Tr}}\left( {\bf Z}_l {\bf{Z}}_l^H  {\bf M}[l,l]\right) - \alpha  \right) =0 $, we have the following two cases:

i) For $\nu=0$, we can attain the optimal ${\bf Z}_l$ as ${\bf Z}_l={\bf X}_{l}+\frac{1}{\rho_2}{\bf D}_{2,l}$,
which must satisfy $ \sum\limits_{l = 1}^L  {\rm{Tr}}\left( {\bf Z}_l {\bf{Z}}_l^H  {\bf M}[l,l]\right) \ge \alpha$. 
  
ii) For $\nu>0$, we have 
\par\noindent
\begin{align}\sum\limits_{l = 1}^L  {\rm{Tr}}\left( {\bf Z}_l {\bf{Z}}_l^H  {\bf M}[l,l]\right) - \alpha  =0,
    \label{53a}
\end{align}
and  the optimal $  {\bf Z}_l $, which is related with $\nu$, as 
       \begin{equation}
  \begin{aligned}
  {\bf Z}_l(\nu)=\left(  \frac{\rho_2}{2}  {\bf I}_{N_{\rm Tx}}- \nu {\bf M}[l,l]  \right)^{-1}\left( \frac{1}{2}{\bf D}_{2,l} +\frac{\rho_{2}}{2} {\bf X}_{l} \right) 
  \end{aligned}\label{53}
  \end{equation}
  
  Let $  {\bf M}[l,l]= \tilde{\bf Q}_l \tilde{\bf \Lambda}_l \tilde{\bf Q}_l^H  $ be the eigen-decomposition of  $  {\bf M}[l,l] $ and 
 $ \tilde{\bf  \Gamma}_l  =   \tilde{\bf Q}_l^H  \left( \frac{1}{2}{\bf D}_{2,l} +\frac{\rho_{2}}{2} {\bf X}_{l} \right)$.
  Thus, $   {\bf Z}_l(\nu) $ can be reexpressed as 
         \begin{equation} 
  \begin{aligned}
  {\bf Z}_l(\nu)=  \tilde{\bf Q}_l \left(  \frac{\rho_2}{2}  {\bf I}_{N_{\rm Tx}}- \nu \tilde{\bf \Lambda}_l \right)^{-1} \tilde{\bf  \Gamma}_l  
  \end{aligned}\label{54}
  \end{equation}
  Plugging \eqref{54} into \eqref{53a} yields 
    \begin{equation}
        \begin{aligned}
  &\sum\limits_{l = 1}^L  {\rm{Tr}}\left(  \left(  \frac{\rho_2}{2}  {\bf I}_{N_{\rm Tx}}- \nu \tilde{\bf \Lambda}_l \right)^{-1}   \tilde{\bf  \Gamma}_l  \tilde{\bf  \Gamma}_l^H     \left(  \frac{\rho_2}{2}  {\bf I}_{N_{\rm Tx}}- \nu \tilde{\bf \Lambda}_l \right)^{-1}  \tilde{\bf \Lambda}_l   \right) \\
  &  = \sum\limits_{l = 1}^L {\sum\limits_{n = 1}^{{N_{{\rm{Tx}}}}} {\frac{{{{{\bf{\tilde \Lambda }}}_l}[n,n]\left( {{{\widetilde {\bf{\Gamma }}}_l}\widetilde {\bf{\Gamma }}_l^H} \right)[n,n]}}{{{{\left( {\frac{{{\rho _2}}}{2} - \nu {{{\bf{\tilde \Lambda }}}_l}[n,n]} \right)}^2}}}} } =\alpha
    \end{aligned}
  \label{55}
  \end{equation}
where $  \left( \tilde{\bf \Gamma}_l \tilde{\bf \Gamma}_l^H\right)  [n,n] $ and $ \tilde{\bf \Lambda}_l[n,n] $ denote  the $ (n,n) $-th element of the matrices $ \tilde{\bf \Gamma}_l \tilde{\bf \Gamma}_l^H  $ and $\tilde{\bf \Lambda}_l $, respectively.
Similar to the solution to \eqref{48}, we can  obtain the optimal solution $\nu^\star$ by utilizing the Newton method \cite{nocedal2006numerical}.
 
\ifCLASSOPTIONcaptionsoff
  \newpage
\fi



%

\footnotesize
\bibliographystyle{IEEEtran}
\bibliography{IEEEabrv,stan_ref}

\end{document}